\documentclass[acmsmall]{acmart}

\setcopyright{acmlicensed}
\acmJournal{PACMMOD}
\acmYear{2023} \acmVolume{1} \acmNumber{N4 (SIGMOD)}
\acmArticle{262} \acmMonth{12} \acmPrice{15.00}
\acmDOI{10.1145/3626756}

\usepackage{booktabs} % For formal tables
\usepackage{graphicx}
\usepackage{balance}
\usepackage{pifont} % for \ding
\usepackage{longtable}
\usepackage{enumitem} % for myitemize and myenumerate
\usepackage[algoruled, linesnumbered, vlined]{algorithm2e}
\usepackage{setspace}
\usepackage{multirow}
\usepackage{url}
\usepackage[T1]{fontenc}
\usepackage{hyperref}
\usepackage{romannum}
\usepackage[compact]{titlesec}
\usepackage{sparklines}
\usepackage[export]{adjustbox}
\usepackage{subfig}
\usepackage{xspace}
\usepackage{xcolor}
\usepackage{soul}
\usepackage[normalem]{ulem}
\usepackage{marginnote}
\usepackage{titlesec}
\usepackage{hyperref}
\usepackage{wrapfig}
\newcommand{\sys}{\textsc{Solo}\xspace}
\newcommand{\sysnostudent}{\textsc{Solo(No student)}\xspace}
\newcommand{\sysnopq}{\textsc{Solo(No PQ)}\xspace}
\newcommand{\sysmle}{\textsc{Solo}\xspace}

\definecolor{coolblack}{rgb}{0.0, 0.18, 0.39}

\newcommand{\tinyskip}{\vspace{3pt}}
\newcommand{\mypar}[1]{\tinyskip\noindent\textbf{#1.}\xspace}

\newcommand{\F}{\mbox{Fig.\hspace{0.25em}}}

\newenvironment{myitemize}{%
\begin{itemize}[leftmargin=1em, itemsep=.1em, parsep=.1em, topsep=.1em,
    partopsep=.1em]}
{\end{itemize}}

\newenvironment{myenumerate}{%
\begin{enumerate}[leftmargin=1em, itemsep=.1em, parsep=.1em, topsep=.1em,
    partopsep=.1em]}
{\end{enumerate}}

\newenvironment{structure*}{\color{blue}\begin{myenumerate}}{\end{myenumerate}}

\sethlcolor{yellow}
\DeclareMathOperator*{\argmax}{argmax}
\DeclareMathOperator*{\argmin}{argmin}

\begin{document}

%%\setlength{\abovedisplayskip}{1pt}
%%\setlength{\belowdisplayskip}{1pt}

\title{Solo: Data Discovery Using Natural Language Questions Via A
Self-Supervised Approach}


\author{Qiming Wang}
\email{qmwang@uchicago.edu}
\orcid{0009-0004-6554-4161}
\affiliation{%
  \institution{The University of Chicago}
  \country{USA}
}

\author{Raul Castro Fernandez}
\email{raulcf@uchicago.edu}
\orcid{}
\affiliation{%
  \institution{The University of Chicago}
  \country{USA}
}



\begin{abstract}

Most deployed data discovery systems, such as Google Datasets, and open data portals only support keyword search. Keyword search is geared towards general audiences but limits the types of queries the systems can answer. We propose a new system that lets users write natural language questions directly. A major barrier to using this \emph{learned data discovery} system is it needs expensive-to-collect training data, thus limiting its utility.

% Data discovery systems help users identify relevant data among large table
% collections. Users express their discovery needs with a program or a set of
% keywords. Users may express complex queries using programs but it requires
% expertise. Keyword search is accessible to a larger audience but limits the
% types of queries supported. An interesting approach is \emph{learned discovery
% systems} which find tables given natural language questions. Unfortunately,
% these systems require a training dataset for each table collection. And because
% collecting training data is expensive, this limits their adoption.

In this paper, we introduce a self-supervised approach to assemble training
datasets and train learned discovery systems without human intervention. It
requires addressing several challenges, including the design of self-supervised
strategies for data discovery, table representation strategies to feed to the
models, and relevance models that work well with the synthetically generated
questions. We combine all the above contributions into a system, \sys, that
solves the problem end to end. The evaluation results demonstrate the new
techniques outperform state-of-the-art approaches on well-known benchmarks. All
in all, the technique is a stepping stone towards building learned discovery
systems. 

%Tabular format is a widespread approach to represent information, and to
%identify relevant tabular datasets among a large collection, traditional  data
%discovery systems provide API or keyword queries which either requires users of
%more techniques or lacks diversity.  Recently, with advances in natural language
%processing, data discovery solutions using natural language question have made
%great progress and are available to a larger audience. 
%
%However, data discovery systems using natural language question need a lot of
%human-annotated training data to get good retrieval accuracy. In addition, for
%each new target table collection, a new set of training data has to be collected
%which is very expensive.  In this paper, we introduce a self-supervised approach
%to generate (synthetic) training data from a target table collection without any
%human involvement.  Further, we propose a new table representation and
%question/table relevance model to implement an end-to-end system S2LD. On two
%well-known benchmark datasets,  S2LD performs close to or even better than
%state-of-the-art systems trained on human-annotated data, and outperforms by
%large margins state-of-the-art systems trained on the same synthetic data. S2LD
%is the first self-supervised data discovery system using natural language
%question and can achieve both good accuracy and little human involvement.

\end{abstract}





\received{April 2023}
\received[revised]{July 2023}
\received[accepted]{August 2023}

\maketitle

\section{Introduction}

The widespread use of tabular formats to represent information, both within
organizations, and across the Internet, means that there is more structured data
available than ever before. This is a boon to both decision makers and citizen
journalists, who can identify alternative data sources to support their
data-driven tasks. Identifying relevant datasets among these large repositories
is challenging due to the large volume of data and the consequent
``needle-in-the-haystack'' problem. In response, several data discovery
solutions~\cite{halevy2016goods, chapman2020dataset, castelo2021auctus,
fernandez2018aurum, zhang2020finding} were proposed over the last few years.
These solutions offer interfaces geared to technical users, such as programmatic
APIs, or keyword search based solutions that are targeted to a more general
audience but that limits the types of queries the system can answer.

\mypar{Example 1/3} \emph{For example, we went to the City of Chicago Open Data
Portal\cite{datacityofchicago} to answer the question ``What
are the business hours of Employment Resource Center at Howard?'', a question that
someone seeking a job in Chicago may be interested in knowing, as employment resource centers help users find jobs. Using the keyword search
interface provided by the portal, we input different variations of our query
such as ``Employment Resource Center'', and ``open hours, employment resource
center'' and more. They yielded different rankings of tables. But none
of them satisfied our information need.}

A promising alternative search interface is to let users express their discovery
needs with \emph{natural language questions}. First, a question lets users express more
precisely the information they need. Second, even non-technical users can ask
natural questions, thus making discovery available to a larger audience. The
huge progress in natural language processing techniques of the last few years
means there are today a few \emph{learned table discovery} approaches that
concentrate in solving exactly this problem: finding a table given a
question written in English. 

\mypar{Example 2/3} \emph{Using \sys, we use its
UI to ask ``What are the business hours of Employment Resource Center at Howard?''
and the system returns a list of tables. The top table named ``Public Technology Resources'' contains the information
we needed, with attributes ``Facility'' and ``Hours'' which do not
exactly match the words in the question but clearly answer our information need.}

Despite their ability to answer natural language questions, a major barrier to using \emph{learned data discovery} systems is they must be trained for each repository before users can search, and collecting training datasets (of question-answer pairs) is difficult, and time-consuming. And because such training dataset must be collected for each new repository where the learned discovery system is to be deployed, and it must be refreshed periodically to account for data changes, this results in a lack of adoption.

In this paper, we introduce a \textbf{self-supervised learned table discovery
approach} that we implement as part of a new discovery system called \sys. \sys implements a new
strategy to automatically assemble high-quality training datasets from repositories of tables without human involvement in neither the collection or labeling of the data, thus tackling the main roadblock to
deploying \emph{learned table discovery} systems. The new strategy relies on self-supervision to construct the training dataset, but we had to carefully design the strategy so it works in practice. In this paper, we concentrate
on table discovery scenarios where the answer to a question $q$ is a table
$T_i^*$ that contains the answer to $q$, e.g., the table ``Public Technology Resources'' in the example
contains the answer to our question. To build \sys we tackled many challenges:

% \begin{myitemize}

\noindent$\bullet$~\textbf{Synthesis of Training Data.} Synthesizing a training dataset off a
repository of tables requires solving two interrelated problems. First, we need
an approach to generate questions from tables to produce positive pairs, $(q,
T_i^*)$, where the table $T_i^*$ contains the answer to the question $q$.
Second, the approach must determine an \emph{appropriate} training dataset size.
Too small, and the system will not generalize. And too large will unnecessarily
consume resources (time and money). Furthermore, choosing a naive heuristic will fail,
as different repositories need different amounts of data. The challenge is to
choose the training data size in a principled way.
% propose a principled approach to choosing the training data. 

\noindent$\bullet$~\textbf{Table Representation for Self-supervision.} The learned discovery systems
ingests tables in vector format. This calls for a table representation strategy.
There are many strategies in the literature to represent tables as vectors, but
we found that many negatively impact performance 
\cite{zhang2018ad, sun2019content, chen2020table, herzig2021open, wang2021retrieving}. 
These strategies either just treat a table as ordinary text and thus ignore the table structure or 
require human-annotation to ensure training data as the same distribution as test data. 
So the challenge is to identify table representation that leads to high performance. 

\noindent$\bullet$~\textbf{Assemble end-to-end system.} Unlike most work in this area, that concentrates on providing algorithms, we provide algorithms and implement them in an end-to-end system. Doing so uncovers several important challenges that if left unaddressed would make the system impractical. We follow a divide-and-conquer approach and split the design and implementation of the system into two broad components, each of which benefits from different machine learning models. During \emph{first-stage retrieval}, the system must identify a subset of
potentially relevant tables. During \emph{second-stage ranking} it must identify
the table that contains the answer to the input question among those returned by
\emph{first-stage retrieval}. When we assembled these two components, we noticed several important performance challenges, such as large storage footprint (due to the distributed representation of many tables), slow encoding time, and low throughput. We implemented a series of techniques to mitigate these problems, achieving several times smaller footprint and encoding time, and subsecond query latency.

% \end{myitemize}

We introduce novel techniques to address the above challenges as part of \sys,
which we will release as open-source. The main contribution of the paper is a
\textbf{self-supervised method} that leverages SQL as an intermediate proxy
between questions and text, facilitating the synthesis of training data. The
approach uses Bayesian neural networks to automatically determine the training
data size in a principled manner. Second, to represent tables in vector format
and feed them to the system, we introduce a simple-to-implement graph-based
table representation, called \textbf{row-wise complete graph}, that is
insensitive to the order of columns or rows, which do not bear any meaning in
the relational model. Third, we introduce a new \textbf{relevance model} that 
%\update{R2O5} {
exploits a pretrained question answering (QA) model \mbox{\cite{izacard2020distilling}} and 
works in tandem with the self-supervised training data collection technique to yield high retrieval performance. Although {\sys} uses pretrained models, users of {\sys} do not need to invest any effort in collecting training data because of {\sys}'s self-supervised training data synthesis.
%} 
Together, these contributions lead to the first \emph{self-supervised learned table discovery} system that automatically assembles training datasets from repositories and lets humans pose their
information needs as natural language questions. 

\begin{figure}
  \centering
  \includegraphics[width=0.65\columnwidth]{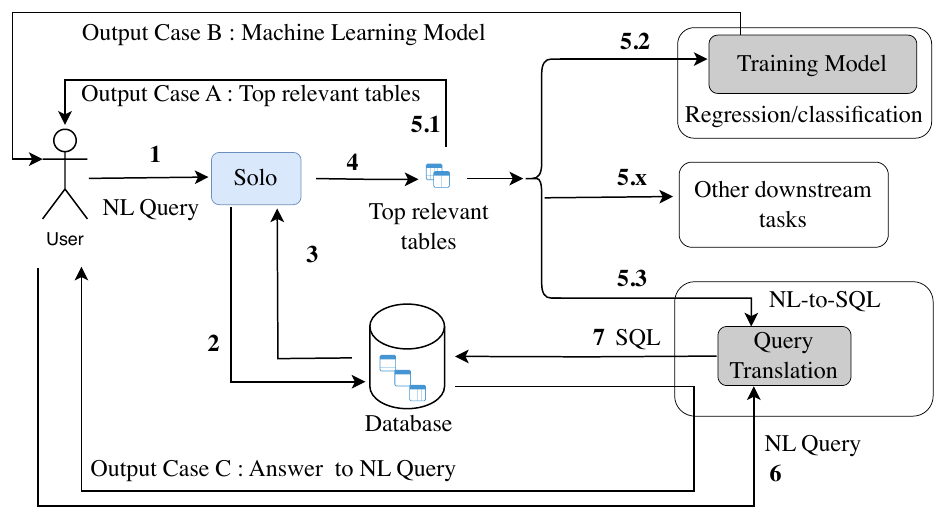}
  \caption{Solo and downstream tasks}
  \label{fig:Solo-Task}
\end{figure}

\mypar{Integrating \sys in Existing Pipelines}
%\update{R1O3, \\ R2O1, \\ R3O1} {
Existing NL-to-SQL interfaces concentrate in answering queries by processing table cells, i.e., query answering \mbox{\cite{hristidis2003efficient, zhong2015answering, popescu2004modern, guo2019towards}}. 
In contrast, {\sys} concentrates on finding a table that is relevant to the user's information needs. Those information needs may be query answering, in which case users can apply existing NL-to-SQL interfaces downstream of {\sys} (see \mbox{\F\ref{fig:Solo-Task}} for an illustrative integration of {\sys} and an NL-to-SQL system). But crucially, those information needs may not be query answering, but training a regression model, or others. Thus, NL-to-SQL systems and {\sys} are complementary approaches, the first concentrating on query answering and the second on data discovery, which is an upstream task.
%}

% {We highlight information needs instead of answer to the question because we don't assume the ultimate goal of users asking natural language (NL) questions. Users may just want to find a table to train a regression model and thus reduce their needs to questions. If users want to find the particular answer cell, it is easy to integrate {\sys} with NL-to-SQL systems \mbox{\cite{hristidis2003efficient, zhong2015answering, popescu2004modern, guo2019towards}} to get the answer as show in
% \mbox{\F \ref{fig:NL-to-SQL}} where NL-to-SQL uses {\sys} to get top relevant tables to translate NL query to SQL which may involve joins.}

We evaluate the ability of \sys to identify relevant tables using existing
state-of-the-art benchmarks and comparing the results with state of the art
approaches---\textsf{OpenDTR} \cite{herzig2021open} and \textsf{GTR}
\cite{wang2021retrieving}.  We show that \sys outperforms other approaches on
previously unseen data repositories, and that it matches and sometimes
outperforms them even when expensive-to-collect training data is provided to the
other baselines, demonstrating the quality of the synthesized training dataset.
We also evaluate the impact of the new row-wise complete graph table
representation, the new relevance model, and the use of Bayesian networks for
efficient training data generation, which may be of independent interest.
Finally, we analyze system-oriented aspects of \sys, including its runtime
performance and reliance on different types of retrieval indexes to give a full
account of the system characteristics and the effectiveness of the optimizations
we implement, which reduces storage footprint by 4x and encoding time by 10x, while yielding sub-second inference times, offering interactive search to users.

Section~\ref{sec:background} presents preliminaries. Section~\ref{sec:solution} presents the new self-supervised strategy, the table representation and relevance model are presented in Section~\ref{sec:trrm}, and \sys in Section~\ref{sec:implementation}. Then, the evaluation is in Section~\ref{sec:evaluation}, followed by 
% related work (Section~\ref{sec:relatedwork}) and 
conclusions in Section~\ref{sec:conclusions}.
 
\section{Preliminaries}
\label{sec:background}

In this section, we present the 
%\update{R1O1 \\ R1O3}{
the scope of data discovery (Section~\mbox{\ref{sec:question_def}})
%}
, and then the problem statement (Section~\ref{sec:problem_def}), followed by the related work (Section~\ref{subsec:discovery_overview}) and their shortcomings, and finishing with the challenges our work tackles in Section~\ref{sec:challenges}.

\subsection{The Scope of Data Discovery}
\label{sec:question_def}

%\update{R1O1 \\ R1O3}{
Our goal is to design a system that takes input natural language (NL) questions and returns tables satisfying users' information needs. NL questions can be exceedingly complicated. Here, we target factual questions \mbox{{\cite{gardner2019question, iyyer2014neural}}}, which are sufficiently expressive to let users discover interesting tables over large repositories. When questions may benefit from several tables, these tables will tend to be higher up in the ranking. Our system's scope is identifying tables relevant to users' information needs without assuming how those tables will be used. For that reason, we do not explore the application of techniques like question decomposition \mbox{{\cite{min2019multi, perez2020unsupervised, zhang2019complex}}} which can be applied to complex questions at inference time to obtain simpler ones that can then be fed to our system. {\sys}, the system we introduce, is a discovery system that sits upstream to the applications users run to solve their tasks. We illustrate this in \mbox{\F\ref{fig:Solo-Task}}, where a NL-to-SQL system is used downstream of {\sys}: {\sys} finds a relevant table given a question, and then the NL-to-SQL system finds the answer to the question within that table.
%}

\subsection{Problem Statement} 
\label{sec:problem_def} 

% \notefb{R3O2. The problem setting is ambiguous. As is illustrated in Section 2.2, it aims at returning a table containing cells that can answer the question. I am afraid this kind of problem setting does not make sense. To the best of my knowledge, existing studies about question answering over tables need to return the cells that are exactly corresponding to the question. Returning a whole table instead of structures in finer granularity provide very little information to the users.}

% \notefb{R3O6. The authors should add more running examples to help readers to understand the problem setting as well as the proposed techniques such as the generation algorithm shown in Algorithm 1.}

A table (relation), $T_i$, has a schema $\mathcal{R}$ with $k$ columns $C_1,
C_2, ..., C_k$ and rows, $r \in T_i$. The table may have a title (caption)
and column names may be missing. A table collection is defined as a set of
tables, $\mathrm{T} = \{T_1, T_2, ... T_N\}$. A cell is indexed by a row $r$ and a
column $c$, and the function $cell_i(r, c)$ returns the content of the cell in
row $r$ and column $c$ of table $T_i$.

Given a question $q$, the problem we solve is to find a table 
$T_i^*$ among a large table collection $\mathrm{T}$, such that $T_i^*$ contains
a cell $cell_i(r, c)$ that answers $q$. This means we do not target more complex question answering scenarios that require combining information from multiple tables.

\mypar{Example 3/3} \emph{The City of Chicago Open Data
Portal has about 630 tables. Today, users input keywords and obtain a ranking of relevant tables. Users can then inspect the table schema and values to determine if the table is indeed helpful to their needs. Our problem statement takes a natural language question as input instead of keywords, and returns a ranking of relevant tables, just as today's search portals such as Socrata \cite{kalinin2022socrat} and Open CKAN\cite{openckan}.}

% We concentrate on returning a list of tables to users, who may want to inspect the tables to build trust on the results before using them. If they are interested in obtaining the specific cells that answer the question, they can plug state-of-the-art table answer models~\cite{herzig2020tapas}, which take 1 table and 1 question as input and return the valid cells. We concentrate on finding the table among many.

\subsection{Related work} 
\label{subsec:discovery_overview} 

\mypar{Table Question Answering} Question answering over tables (Table QA) models like \textsf{TAPAS}~\cite{herzig2020tapas}, \textsf{Tabert}~\cite{yin2020tabert} and \textsf{Tapex}~\cite{liu2021tapex} take 1 question and 1 table as input and return the cell that answers the question. Our system is complementary as it takes a \emph{collection of tables} and returns the table necessary to answer the question. Both approaches can be combined if one wishes to directly obtain the answer for the question, although in practice analysts want to understand the table from where the answer is obtained to build trust on the result. 

\mypar{Table Search} The ad-hoc table search task in information retrieval 
\cite{zhang2018ad, sun2019content, chen2020table} also tries to solve data discovery problems 
but focus on key-word search which often have many relevant tables  
while our task needs to identify the table that can answer an input question and so needs more accurate relevance modeling. 

\mypar{Learned Data Discovery Systems} More recently, \textsf{GTR} \cite{wang2021retrieving} uses graph representation learning to model the relevance between a question and a table, retrieving the table with the highest relevant to a question. However, this model does not scale to a large collection of tables because computing the relevance is expensive and must be done for each table. Our contribution, \sys, first retrieves a small collection of relevant tables, and then applies a relevant model on this subset. A relevant approach is \textsf{OpenDTR} \cite{herzig2021open} which learns a dense question and table encoder for large scale table search. However, \textsf{OpenDTR} requires expensive human-annotated data to perform well, so do the non-scalable relevance models.

\mypar{Natural Language to SQL (NL-to-SQL)} 
%\update{R1O3 \\ R2O1 \\ R2O7 \\ R3O1 \\ R3O2}{
NL-to-SQL approaches aim to provide a NL interface to databases \mbox{\cite{hristidis2003efficient, zhong2015answering, popescu2004modern, guo2019towards}} for query answering. To this end, they translate a NL query to a SQL statement which is executed by a SQL engine to get the answer to the query. One of the challenging subtasks of a NL-to-SQL system is schema matching, necessary to map NL to valid table and column names \mbox{\cite{katsogiannis2023survey}}. In existing NL-to-SQL benchmarks, datasets often contain a small number of tables and also each query often includes the table names which makes finding the right tables trivial, e.g. there are only 5 tables on average per each database in the impactful Spider \mbox{\cite{yu2018spider}} dataset. In contrast, data discovery systems such as {\sys} aim to find a relevant table among tens of thousands of tables. In summary, NL-to-SQL approaches concentrate in solving query answering, while {\sys} concentrates in addressing data discovery.
%}

\mypar{Large Language Models (LLM)}
%\update{R1O2 \\ R3O2}{
LLMs such as GPT-4 \mbox{\cite{gpt4}} answer input questions from a static collection of data, but, despite recent progress in that direction, it is harder to use them to answer questions over newly incorporated data because retraining them is prohibitively expensive. In addition, they are incapable of indicating what input tables (or data) were used to answer a prompt, which is necessary in discovery scenarios so users understand the provenance of the results. {\sys} adapts quickly to new repositories using the new self-supervised technique, and its answers exactly determine what tables contain the answer to the input question.
%}

\mypar{Retrieval-based question answering over documents (RQA)}
This class of systems concentrate on query answering over textual documents, as opposed to relational data. RQA approaches often include two components, a \emph{retrieval} and a \emph{reader} \mbox{\cite{chen2017reading}}. The retrieval gets top k documents from large repositories so that the reader can output answers quickly \mbox{\cite{karpukhin2020dense}}. Nowadays the reader is often implemented by a LLM and much work is about how to combine the retrieval and LLM for best performance \mbox{\cite{izacard2020leveraging, lewis2020retrieval, borgeaud2022improving, mialon2023augmented}}. For example, RAG \mbox{\cite{lewis2020retrieval}} uses a dense document retriever to augment a seq2seq language model and trains them jointly to output the best answer. But RQA concentrates on documents and not tables. Applying RQA to relational data discovery requires addressing the challenges of table representation (as documents) and table (document) ranking which are not trivial if human-annotated (question, table) pairs are unavailable. These are precisely the challenges our work addresses.

\mypar{Scaling laws for neural models} 
%\update{R2A1 \\ R2O3}{
The performance of neural models largely depends on three factors\mbox{\cite{kaplan2020scaling}}: model size,  training data size and compute budget. A lot of work \mbox{\cite{kaplan2020scaling, ghorbani2021scaling, alabdulmohsin2022revisiting, bahri2021explaining, cherti2023reproducible}} has shown that there is a power-law relationship between the performance and any of three factors. That means increasing model size, training data size or compute budget results in diminishing returns in performance. We observe similar trends during training, even though it remains difficult to write a concrete formula for determining when to stop training. Instead, to save compute budget, we combine Bayesian incremental training and self-supervised approach to automatically determine the appropriate training dataset size as described in Section~{\ref{subsec:bnn_train}.}
%}

\subsection{Challenges of Learned Discovery Systems}
\label{sec:challenges}

Existing learned discovery systems suffer from several shortcomings that
delineate the challenges we tackle in this paper.

\mypar{C1. Collecting training data} Learned discovery systems trained on a
table collection do not perform well in new table collections, so they have to
be freshly trained for each table collection. Collecting training data is
expensive and a major limitation of learned discovery systems. 
For example, \textsf{OpenDTR} uses 12K <question,table-with-answer> pairs. The
challenge is to avoid the manually intensive task of collecting and labeling
training data and deciding how large the training dataset should be to become
representative and thus lead to a system that performs well on the target table
collection.

%Learned discovery systems rely on the existence
%of large training datasets. 
%For example, \textsf{OpenDTR} uses 12K
%<question,table-with-answer> pairs. 
%A model trained in one table collection will
%underperform in a different table collection, as we show in the evaluation. The
%challenge is to avoid the manually intensive task of collecting and labeling
%training data.

%To train the question encoder and table encoder
%for the first-stage retrieval  or semantic relevance model for the second-stage
%rank, a user has to collect many (question, right table) pairs.  E.g. In OpenDTR
%\cite{herzig2021open}, 12K examples are used to train the two encoders.  To make
%it worse, a model trained on one table collection often performs badly on
%another one. That means,  to achieve good performance, a user has to repeat
%collection of training data  for every new domain table collection. This is
%tedious and involves large human cost. 

\mypar{C2. Table Representation} Learned discovery systems represent tables as
vectors. \textsf{OpenDTR} uses \textsf{TAPAS}~\cite{herzig2020tapas} to represent a table
with a single vector. As we show in the evaluation, the sparse and simpler BM25
model sometimes outperforms TAPAS. 
%Other models, such as
%\textsf{MTR}~\cite{shraga2020web} represent tables using features such as the
%name, column names, row cell records and what they call \emph{facets}, which are
%vertically divided cell records. 
More recently, \textsf{GTR}~\cite{wang2021retrieving} represents tables as
graphs, similar to other efforts in data managements such as
\textsf{EmBDI}~\cite{cappuzzo2021embdi} and Leva~\cite{zhao2022leva}. However,
the graph construction depends on the order of rows and columns, which has no
meaning in relations. Table representation has a big impact on system
performance. Thus, the challenge is to find a dense table representation that
outperforms sparse models and that works well on relations (\textbf{C3}).

%\mypar{Table representation \& relevance model} A table contains multiple
%fields, the title (caption), column headers, and cell data with many rows and a
%question can refer to any part of the table.  A good table and relevance
%representation are very important to achieve good performance for first-stage
%retrieval and second-stage rank respectively.  

%For first-stage retrieval, OpenDTR \cite{herzig2021open} uses two instances
%(different parameters) of the Tapas \cite{herzig2020tapas} model  to implement
%the question and table encoder respectively. Tapas is first proposed as a table
%question answering (Table QA) model which takes a question and table as input
%and return some cells of the table and an aggregation oprator as the answer.  In
%OpenDTR, the question Tapas model  takes a question only as input, while in the
%table Tapas model, the table title is the first input and the remaining of the
%table is the second input. While Tapas is a good Table QA model, our experiment
%shows, OpenDTR performs much worse than  BM25 on some dataset in finding the
%right table. It seems questionable to repesent a whole table using only one
%vector. 

\mypar{C3. Relevance Model} The model computes a question representation and
measures the relevance of different tables with respect to the question. The
model must be designed in tandem with table representation (\textbf{C2}). The
challenge is to find a relevance model appropriate for relations that performs
well empirically when trained with synthetically generated questions (\textbf{C1}).

%MTR\cite{shraga2020web} is a state-of-the-art relevance model for the
%second-stage rank. It treats a table as a multimodal object and each modal
%represents one of the 4 components, the table Title, column headers (names),
%cell records and facets (vertically divided cell records and one for each
%column). Relevance representation between the question and each modal is
%computed and then a gating machanism is applied to get a combined relevance
%representation between the question and table. Recently,
%GTR\cite{wang2021retrieving} represents a table as a graph.  In the graph,
%neighboring (in row/column) cells are inter-connected and there are global
%column and row nodes for each column and row repesctrively and a directed edge
%from each cell node to its corresponding column and row nodes.  A Graph
%Transfomer network is then applied to learn the table and relevance
%representation. We argue that neighter MTR or GTR captures the essential
%relevance of a question and table. i.e. relation match. and we introduce our
%solution in next section.  

\section{Self-supervised Training Data Collection}
\label{sec:solution}

In this section, we present the new self-supervised strategy for training learned discovery systems, thus tackling challenge \textbf{C1}. Creating a training dataset requires solving two problems; we introduce novel strategies to address them:

\begin{myitemize}
\item \textbf{Generate question - table pairs}. We introduce a strategy to automatically generate pairs of questions and the tables with the answers (Section~\ref{subsec:syt_data_generation}).
\item \textbf{Determine training dataset size} We introduce a strategy to determine when the size of the training dataset is sufficient to train the system (Section~\ref{subsec:bnn_train}).
\end{myitemize}

Throughout this section, we assume the existence of two additional components, a table representation component and a semantic relevant model component, that are needed to create the training dataset. Here, we treat these two components as black boxes. Then, in the next section, we provide the details of their design and implementation.

% First, we present the new self-supervised training method to automatically create 
% synthetic labeled questions without manual intervention (Section~\ref{subsec:syt_data_generation}).
% Next, we introduce a new table representation that boosts the performance of
% first-stage retrieval
% (Section~\ref{subsec:table_repre}). Finally, we present the
% relevance model that implements the second-stage ranking and uses the
% automatically created data to complete our contribution
% (Section~\ref{subsec:relevance_model}). \notedel{ In Section~\ref{subsec:bnn_train}, 
% we present Bayesian incremental training to automatically choose the appropriate data size and 
% train the  relevance model. We conclude by presenting \sys in 
% Section~\ref{sec:implementation}}

\subsection{Synthetic Question Generation}
\label{subsec:syt_data_generation}

A training dataset consists of positive and negative samples of question-table pairs. We first explain how to generate positive samples. We explain how to collect negative samples in Section~\ref{subsec:negative}.
% Here, we explain
% the method to generate synthetic questions and detail how to derive a training
% dataset from these in Section~\ref{subsec:relevance_model} and Section~\ref{subsec:bnn_train} , thus completing the
% solution to \textbf{Challenge 1}.

The goal is to obtain a collection of question and table pairs, <$q, T_i^*$>,
where $q$ can be answered with $T_i^*$. To automatically generate natural
questions from tables a simple, yet problematic, approach is to define question templates and then fill in the templates with data from the tables. Unfortunately, this approach does not lead to diverse questions, which are needed to increase the matching of questions written by users, who may write questions very different than the templates.

Instead, we exploit the following observation. Large language models excel at giving diverse natural language interpretations of a SQL query, so we first generate SQL queries from tables and then translate those SQL queries to natural language. Specifically, our strategy consists of 3 different stages that we explain next: i) design a SQL template structure; ii) generate SQL queries from tables; iii) translate SQL queries to natural language.

\subsubsection{SQL Template Structure.}

We generate SQL queries that match the structure of \emph{factual questions}; those answered with one table and that do not require complex processing and reasoning. More precisely, given a table $T_i$ that represents an entity $E_i$ with attributes $E_i.A_1$ , … ,  $E_i.A_{m_i}$, the answer to a factual question exists in some $E_i.A_j$ whether raw or via an aggregation operator such as $(Max, Min, Avg, Count)$. Factual questions are those that can be answered by search engines using a knowledge base~\cite{west2014knowledge}. We use this format to drive the design of SQL templates, which are generated by applying the following rules:

% More precisely, given a table $T_i$ that represents an entity $E_i$ with attributes
% $E_i.A_1$ , … ,  $E_i.A_{m_i}$, the answer to a factual question exists in some
% $E_i.A_j$ whether raw or via an aggregation operator such as $(Max, Min, Avg,
% Count)$. The factual question may optionally incorporate operators that further
% qualify the scope of the answer, such as $(>, <, =)$ (operators apply only to
% compatible data types).

% As a result, the SQL query that represents a factual question tends to be simple,
% with attributes and (maybe) aggregation functions and predicates that use the
% operators above. These questions are helpful for table discovery as evidenced by
% the format of popular benchmarks such as \textsf{NQ-Tables} that consist of real
% queries from Google users and which we use in the evaluation section. Consequently, we use this format to drive the design of SQL templates, which are generated by applying the following rules:

\begin{myitemize}
\item Randomly include an aggregation operator, \texttt{MAX, MIN, COUNT, SUM,
AVG} on numerical columns.
\item Include $\geq 1$ columns with predicates of the form ``= < > ''.
\item Do not include joins as we seek a single table that answers $q$.
\end{myitemize}

In generating questions we strive for including sufficient \emph{context} and
for making the questions sufficiently \emph{diverse} to resemble questions posed
by real users. In more detail:

\mypar{Context-rich queries} Whenever available, we include the table title with
certain probability (as detailed below) as a special attribute (column)
\emph{``About''} to generate questions. This is because the table title often
incorporates useful context. Consider the question ``Where was he on February
19, 2009?''. This cannot be answered because we do not know what ``where''
refers to and who is ``he''.  The question, however, makes more sense if a table
titled ``List of International Presidential trips made by Barack Obama'' can be
integrated. 
%The resulting question does not directly copy the title and instead
%incorporates tokens that correspond to named entities. 
We do not always include the title because it causes the learning procedure to overemphasize its
importance and ignore the table schema. Instead, we introduce the title with
certain probability $\alpha$. If a SQL query contains $m$ attributes in the
predicates, we include the title with probability $\alpha = 1 / (m+1)$: more
attributes means more context is already available so the title is less likely
helpful.

\mypar{Diverse queries} We want synthetic questions to resemble those posed by
users.  One way to achieve that is to ensure the questions represent well the
attributes contained in the tables. To achieve that, we control how many
predicates (referring to different attributes) to include in each question by
sampling at most $m$ attributes. In addition, if the ``About''
attribute from above is included, it makes a question refer to as many as $(m+1)$ attributes.

\subsubsection{Generation Algorithm}

The generation algorithm must: i) deal with dirty tables; ii) sample to avoid
generating humongous training datasets when the number of tables in the input is
large. We first explain the strategy to deal with these two challenges and then
present the algorithm.

\mypar{Dealing with dirty data} Dirty tables in our context are those that become hard to generate SQLs from. They include those with missing column names and cell values with too long text. They are common in open data repositories~\cite{cafarella2008webtables, herzig2021open} and even enterprise scenarios.
% that may result from incorrect parsing (such as when tables are automatically crawled from the web, as in webtables \cite{cafarella2008webtables} and NQ-Tables \cite{herzig2021open}.  
A cell with too long text is a problem when translating from SQLs to questions because the translation model we are using allows 200-300 words at most as input. 
Cells with text longer than $Q3 + 1.5 (Q3 - Q1)$, where $Q1$ and $Q3$ are the first and third quartiles of the distribution of text length, are discarded.
% IQR is a simple yet robust way of detecting outliers \cite{peter2021srm}. 
When generating SQLs, the algorithm ignores columns without header names because the translation model asks for a relationship as input and the column name is the only option we have. Alternatively we can recover the headers using similarity search \cite{harmouch2021relational}.

%We found some tables in the \textsf{NQ-Tables} \cite{herzig2021open} dataset are dirty, e.g.
%1) missing column headers,  
%2) nonsense  column headers,  
%3) column headers and cell text are the same,  
%4) incorrectly aligned rows, i.e. some row is shifted by one column, 
%5) long and duplicate cell text.
%SQLs sampled from these dirty tables may result in bad questions and thus mislead the model.  
%We assume at test time, people will write good questions so apply some strategies to remove those bad questions. 
%We assume bad questions are too hard or too easy to answer and so this reduces to detecting easy and hard training samples. One approach to measure %easiness and hardness of learning from a training example is to compute the “forgetting” times during training. Specifically, we train an off-the-shelf Table QA %model using the generated questions. During training, after some batch steps,  we use the checkpoint model to predict the output of an input example. After %some training steps, we then summarize how often an example switches from correct prediction to incorrect (forgetting) . Intuitively,  an example that always %gives correct predictions or has very few forgettings is easy to learn, while an example that always gives incorrect predictions  is hard to learn.

\newcommand\mycommfont[1]{\footnotesize\ttfamily\textcolor{blue}{#1}}
\SetCommentSty{mycommfont}
\SetKwComment{Comment}{//}{}

\begin{algorithm}
\small
\caption{$generate\_sqls(T, batch\_size, sql\_dict)$} \label{alg:sql_generation}
$SQLs = []$ \\
\While{len(SQLs) < batch\_size}{ 
	$T_i = sample\_table(T)$ \\
	$sel\_col = sample\_columns(T_i.columns, 1)$  \\
	$num\_cond = sample([0, .. , Max\_Cond\_Num])$ \\
	$cond\_cols = sample\_columns(T_i .columns, num\_cond)$  \\
	$\alpha = 1 / (len(cond\_cols) + 1)$ \label{alg:line:alpha} \\
	$use\_title = bernoulli.rvs (\alpha, 1) $ \\
	\If {$use\_title$} {
		$cond\_cols.append(TITLE)$ 
	}
	$agg\_op = sample\_aggre\_operator()$ \label{alg:line:sample_aggr} \\
	$good\_rows = collect\_rows(T_i, cond\_cols)$  \\ 
	\If {$good\_rows$ is not empty} {
		$r = sample\_row(good\_rows)$ \\
		$sql = create\_sql(r, sel\_col, cond\_cols, agg\_op)$ \label{alg:line:create_sql}\\
		\If {sql not in $sql\_dict$} {
		     $SQLs.append(sql)$ ; $sql\_dict.add(sql)$
		 }
	 }
}
\Return SQLs
\end{algorithm}

\mypar{Sampling strategy} 
%\update{R2O3}{
We do not know how large the training dataset must be to achieve good performance. And extracting all possible SQL queries from a table collection would lead to too large datasets and infeasible training times.
%} 
For example,
consider a table collection with 100 tables, where each table has 10 columns and
only 12 rows, we sample 4 columns per table and use only one
aggregator operator. In this case, the number of unique SQLs is at least 
$100 \cdot 12 \cdot 10 \cdot \left(\sum_{k=0}^{3}{{10}\choose{k}}\right)  = 2,112,000$. 
%And that is just the number of positive samples. 
This number grows fast with more tables and rows. Training on so many questions
requires more hardware resources and time and quickly becomes prohibitively
expensive. 

Instead, the generation algorithm
uses a sampling procedure that is driven by the training process which asks for
a collection of $batch\_size$ SQLs to be generated. This procedure is used by the incremental training strategy we introduce (Section~\ref{subsec:bnn_train}) to determine when a training dataset is sufficiently large to achieve good performance on the task and thus stop. Now, within each invocation of the algorithm, we want to ensure that the $batch\_size$ SQLs generated cover the table collection well. We achieve that by using the following algorithm.

\noindent\textbf{The Generation Algorithm.} The algorithm combines 3
independent sampling procedures (Algorithm~\ref{alg:sql_generation}). The first
samples tables. The second samples columns given a table, and includes one of
the logical operators in the predicates. Then the $alpha$ probability is
computed (line~\ref{alg:line:alpha}) to decide whether to include the table's title. The third
sampling procedure samples rows, given columns of a table. When columns are numerical,
the generation algorithm samples one of the aggregate operators. In addition, and ahead of the sampling process, 
the algorithm makes an initial pass over the data to: i) determine column
types; and ii) compute the distribution of cell text length, which is used to filter out outliers. The
result of the algorithm is some number ($batch\_size$) of SQL queries used for training or validation.

\subsubsection{Translation from SQL to question.}

We use LLMs to translate SQL queries into natural language questions. Specifically, we use the
T5~\cite{raffel2019exploring} sequence-to-sequence model. T5 is
pretrained on the ``Colossal Clean Crawled Corpus'' (C4) \cite{dodge2021documenting}, 
which includes internet documents from 365 million domains with 156 billion tokens. 
Although C4 is very large, it is mostly in natural language, and the pretrained T5 model 
learns a natural language to natural language mapping which is not directly applicable  for our purpose. So we \emph{fine-tune} T5 using
WikiSQL~\cite{zhongSeq2SQL2017}, a dataset with pairs of SQL queries and their
corresponding natural language representation that has been previously used for
tasks such as natural language question interface to relational databases ~\cite{zhongSeq2SQL2017}.

\begin{table}[h]
\resizebox{0.65\columnwidth}{!}{%
\begin{tabular}{|l|l|l|l|}
\hline
\textbf{SQL keyword} & \textbf{Custom T5 token} & \textbf{SQL keyword} & \textbf{Custom T5 token} \\ \hline
select               & {[}S-E-L-E-C-T{]}   & where               & {[}W-H-E-R-E{]}     \\ \hline
avg                  & {[}A-V-G{]}         & =                   & {[}E-Q{]}           \\ \hline
max                  & {[}M-A-X{]}         & \textgreater{}      & {[}G-T{]}           \\ \hline
min                  & {[}M-I-N{]}         & \textless{}         & {[}L-T{]}           \\ \hline
count                & {[}C-O-U-N-T{]}     & and                 & {[}A-N-D{]}         \\ \hline
sum                  & {[}S-U-M{]}         &                     &                     \\ \hline
\end{tabular}%
}
\caption{SQL keywords for T5 model input}
\label{tab:t5-SQL}
\end{table}

During fine-tuning, we update the weights of T5 without adding new layers. When
encoding SQL statements, and to escape SQL keywords, we include these in
brackets following a format that indicates the model they must be escaped, e.g.,
``where'' becomes ``[W-H-E-R-E]''. A complete list of keywords is shown in table \ref{tab:t5-SQL}; those keywords will never appear in the output questions.

%A training table, SQL example and
%corresponding question is showsn in \F\ref{fig:sql2question}. Special tags like
%[S-E-L-E-C-T] are used to annoate SQLs to avoid possbile collison with tokens in
%tables.  

%\begin{figure}[t] 
%\centering
%\includegraphics[width=\columnwidth, trim={4.5cm, 3.5cm, 5.5cm, 1.5cm}, clip]{img/sql2question}
%\caption{An example of SQL and corresponding question from a table} 
%\label{fig:sql2question} 
%\end{figure}

\subsubsection{Assigning $T_i^*$.}

Large table collections contain table duplicates and near-duplicates. 
Consequently, a single question may be answered
by more than a table. The approach considers
\emph{any} table
that can answer a question as a valid solution, and thus it creates multiple
question table pairs. To detect duplicates, our
approach simply filters out tables with the same schema; it can be extended to use duplicate detection techniques~\cite{fernandez2018aurum,koch2023duplicate} .

\subsubsection{Validation data.}
It is a best practice to sample validation data
from the same distribution as the test data. However, this requires users to manually collect
question and answer table pairs. 
%\update{R2O2}{
Our goal is to not require manual labeling by users, which we achieve via the new self-supervised method. Now, we collect validation data in the same way we collect training data, using a holdout set that the system automatically creates on the input table collection.
%}

% and so reserve 20\% tables and generate 2,000 synthetic questions from them for validation.} 
% Following this method, our approach resembles table pre-training and zero-shot learning \cite{wang2019survey} 
% where the pre-training task is similar to or the same as the downstream(test) task but the data used 
% for pre-training are not i.i.d examples from the downstream(test) data distribution. 

%\notera{this very last sentence needs more explanation}

\subsubsection{Completing training dataset}\label{subsec:negative} The discussion so far has concentrated on generating positive samples of question-table pairs. Now we explain how to generate negative samples. The data is collected to train the relevance model that we explain in detail in Section~\ref{subsec:relevance_model}. During test, the relevance model always takes as input a small set of tables from first-stage retrieval. 
To emulate the test scenario, given a synthetic question,  the system calls the first-stage retrieval component to get a small set of tables and labels those tables not the source of the question as negative examples which are often called hard negatives \cite{schroff2015facenet}. If all tables are positive or negative we ignore this question. The first-stage retrieval does not retrieve all the cells for a table but some cells of it because of the table representation which we explain more in Section~\ref{subsec:table_repre}.

\subsection{Bayesian Incremental Training}
\label{subsec:bnn_train}

We introduce a strategy that increases the training dataset size incrementally until achieving good performance, thus, removing the need to select a training dataset size a priori.

% A naive way of solving the problem is to successively generate training
% datasets of increasing size, retrain the model, and stop whenever the quality
% is satisfactory. The shortcoming of this approach is that each training process
% is independent from the previous and takes larger and larger inputs, hence,
% making it more and more time consuming. We need a better strategy.

Our idea is to apply a new Bayesian incremental training
process~\cite{kochurov2018bayesian}. We first provide technical background on Bayesian Neural Networks and then present our incremental strategy.

\subsubsection{Background on Bayesian neural networks.}
\label{sec:bnn} 

Bayesian neural networks attach a
probability distribution to the model's parameters, thus paving the way
to avoid overfitting and providing a measure of uncertainty with the answer.
We leverage Bayesian neural networks for a different purpose, to learn when a
model's performance has reached a good quality and thus avoid keep feeding
expensive to generate training data. 
% In other words, they are the workhorse
% behind the technique we use to implement incremental training.

A general approach to learn the parameters $\theta$ of a
neural network from dataset $D$ is to maximize the likelihood,
$\tilde{\theta} = \argmax {P(D|\theta)}$. This point estimation of the
parameters often leads to overfitting and overconfidence in predictions
\cite{jospin2022hands}. To address this issue, Bayesian neural networks
\cite{jospin2022hands, blundell2015weight} learn a posterior
distribution over $\theta$ using the Bayes rule. 
\begin{align*}
P(\theta|D) = \frac {P(D|\theta)P(\theta)} {P(D)} = \frac {P(D|\theta)P(\theta)} {\int_{\theta} P(D|\theta)P(\theta) d\theta} 
\end{align*}

At inference time, the prediction is given by taking expectation over $P(\theta|D)$, 
\begin{align*}
P(y|x) = \mathbb{E}_{P(\theta|D)} (P(y|x, \theta))
\end{align*}
In practice, a list of $\theta_1, \cdots , \theta_m$  are
sampled from $P(\theta|D)$ and the prediction is made using bayesian model
averaging \cite{wilson2020bayesian},
\begin{align*}
P(y|x) = \frac{1} {m} \sum_{i=1}^{m} P(y|x, \theta_i)
\end{align*}

While it is easy to define a network for $P(D|\theta)$, such as the multilayer
perceptron (MLP), and specify a prior for $\theta$, such as an isotropic
gaussian, the posterior distribution $P(\theta|D)$ is computationally
intractable because of the integration over $\theta$. To address the
intractability, variational learning \cite{blei2017variational} uses another
distribution $q(\theta|\varphi)$, parameterized by $\varphi$ to approximate
$P(\theta|D)$.  Variational learning optimizes $\varphi$ to minimize the KL
divergence of $P(\theta|D)$ and $q(\theta|\varphi)$. We use Bayesian neural
network to train the relevance model given a list of datasets $D_1, ... D_m$.

\subsubsection{Bayesian Incremental Training Strategy.}
The goal is to learn recursively a
posterior distribution $P(\theta|D_1, ... , D_t)$ over the neural network
parameters $\theta$ given a prior distribution  $P(\theta|D_1, ... , D_{t-1})$
and only one dataset $D_t$.
% , instead of an accumulation of datasets, such as in
% the simple baseline explained above. 
The prior distribution $P(\theta|D_1, ... ,
D_{t-1})$ contains the knowledge the relevance model learns from previous
datasets so that the model does not have to start training from scratch and uses
$D_t$ only. This is the key to the efficiency gain.

Because the posterior
distribution $P(\theta|D_1, ... , D_t)$ is intractable to compute, we
use $q(\theta|\varphi)$, parameterized by $\varphi$ to
approximate the posterior and then optimize $\varphi$ using backpropagation \cite{blundell2015weight},
\begin{align*}
\tilde { \varphi } = \argmin { \sum_{i=1}^{m} \log {q \left (\theta^{i}|\varphi \right)} - \log {P \left (\theta^{i} \right )} - \log {P \left (D|\theta^{i} \right )} } 
\end{align*}

Where $\theta^{i}$ is a sample from $q(\theta|\varphi)$ and $P \left
(D|\theta^{i} \right )$ is determined by the relevance model (see Section
\ref{subsec:relevance_model}). Next, we discuss the choice of
$q(\theta|\varphi)$ and $P (\theta)$ and also some techniques to speed up training.

\mypar{The approximate distribution, $q(\theta|\varphi)$ } In the relevance
model (see next section) $\theta$ = \{($\boldsymbol{W_i}$ , $\boldsymbol{b_i}$)\}, where
$\boldsymbol{W_i}$ and $\boldsymbol{b_i}$ are the weight matrix and the bias
vector for layer $i$. To simplify and speedup computation, we
follow the approach in \cite{blundell2015weight} and assume that each weight and bias
is an independent gaussian variable with mean $\mu$ and standard
deviation ${\sigma}$, so that $q(\theta|\varphi)$ can be fully factorized.
More concretely, random noise $\epsilon$ is sampled from the unit gaussian
$\mathcal{N}(0, 1)$, and  ${\sigma}$ is derived by $\sigma = \log (1 +
\exp(\rho))$ and then a sample weight $w$ in $\boldsymbol{W_i}$ is given by $w
= \mu + \sigma \cdot \epsilon = \mu + \log (1 + \exp(\rho)) \cdot \epsilon $.
So the parameter $\varphi$ is the set of $(\mu_j, \rho_j)$.

To initialize $\mu_j$, we sample uniformly from $(-1, 1) $.  To initialize
$\rho_j$, we sample uniformly from $(-3, 0)$ and this makes the initial
$\theta_j$ sit in the range $(0.05, 0.7)$. Smaller initial $\theta_j$ for each
weight/bias would make training much harder to converge.

\mypar{The prior $P (\theta)$ } We only have to specify a prior for the first
dataset $D_1$. After training is done with $D_1$, the posterior $P(\theta|D_1)$
is the prior for the next training and so on. Then both the prior and posterior
are gaussian distributions. In particular, the initial prior is a gaussian with mean
$0$ and standard deviation $1$ for each weight or bias, which is 
consistent with the initialization of the parameter $(\mu_j, \rho_j)$. In
addition, in our scenario, an inaccurate initial prior is not an issue because there
are many training data points and as the training data increases, the impact of the
initial prior plays a lesser role.

\mypar{Speedup of training and evaluation} 
First, we train the Bayesian neural network by sampling $\epsilon$ once per batch (as opposed to $n$ times as in other context), thus speeding up the training process without losing accuracy. Sampling $n$ times is usually done to conduct model averaging: we achieve the same effect by sampling across iterations.
% To train a Bayesian
% neural network, often multiple random noise ($\epsilon$) are sampled for each
% batch and during evaluation also multiple noises are sampled to do bayesian
% model averaging. This, however, would make the Bayesian
% neural network training more time consuming than the baseline procedure, even
% when the baseline procedure trains over a larger training dataset.
% To address the issue, we use one sample of random noise only during
% training; we find that there is no performance degradation for our model. 
Second, on validation data, 
we use the posterior gaussian mean of weights/bias for prediction instead of doing
bayesian model averaging, further reducing time. Together, these optimizations
improve the training time of the Bayesian neural network, making it
substantially faster than the baseline approach.

\section{Table Representation and Relevance Model}
\label{sec:trrm}

In this section, we present the strategy for representing tables as vectors (Section~\ref{subsec:table_repre}) and the semantic relevance model (Section~\ref{subsec:relevance_model}), tackling challenges \textbf{C2} and \textbf{C3}.

\subsection{Table Representation}
\label{subsec:table_repre}

% \notefb{R3M2. The reference [44] also models a collection of tables as a graph to learn richer information. Could the authors explain the difference between this paper and Section 4.2 of this paper? The current statements in Section 2.4 is too high level and didn't explain the advantage of new techniques in Section 4.2 over those in [44].}

% \notefb{R2O3. The paper could expand upon its work in triplet generation and the row-wise complete graph. I liked the paper's ideas here and think the paper would benefit from expanding the textual description here to be more thorough, and by conceptually describing in more detail how this improves upon prior representations of tables.}

The goal is to find a table representation that facilitates matching with
questions, thus addressing \textbf{Challenge 2 (C2)}.

We represent each table as a graph, where the nodes are the \texttt{subject} and
\texttt{object} of every (\texttt{subject, predicate, object}) triple in the
table. The intuition is that a natural question that can be answered with a
table can be answered with a collection of triples from that table.  Hence,
representing the table via its triples facilitates matching it with relevant
questions during first-stage retrieval. There are two advantages to representing tables as triples. First, a triple represents a basic relationship. By representing tables with its triples, we obtain a more fine-grained representation than state of the art models such as \textsf{OpenDTR}, which treats a table as a whole. Second, it is easier to treat a triple as text than it is to treat a table, so we can use any off-the-shelf text encoders to transform triples into their vector representation. Next, we provide an example first, present the new \textbf{row-wise complete graph} method for table representation, and finish with encoding.

\begin{figure}[t] 
\centering
\includegraphics[width=0.65\columnwidth]{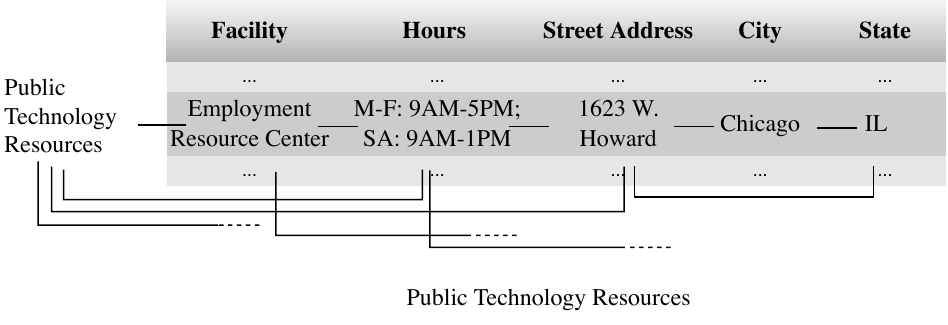}
\caption{Example of Row Wise Complete Graph (\textsf{RCG})} 
\label{fig:table_repre_example} 
\end{figure}

\mypar{Example} Consider the running example 
"\textit{What are the business hours of Employment Resource Center at Howard?}", and 
the table that answers this question, $T_i^*$, shown in \F~\ref{fig:table_repre_example}. The question contains two triples:
\textit{(Employment Resource Center,  at,  Howard)} and \textit{(Employment Resource Center,
business hours,  ?)} , where the placeholder ``?'' corresponds to the answer.  At
the same time, the table contains two triples related to the question:

\begin{myitemize}
\item \textit{ (Employment Resource Center , Facility - Street Address ,  1623 W. Howard)} denoting the
relationship between \emph{Facility} and the \emph{Street Address} column, i.e. Employment Resource Center is located at Howard street, 
which is close to \textit{(Employment Resource Center,  at,  Howard)} in the question. 

\item \textit{ (Employment Resource Center , Facility - Hours , M-F: 9AM-5PM; SA: 9AM-1PM)} 
denoting the relationship between \emph{Facility} and \emph{Hours} columns, with two column names connected by a hyphen is the
predicate. This relationship is close to \textit{(Employment Resource Center ,  business hours ,  ?)} implied in the question. 

\end{myitemize}

Matching triples is the mechanism our approach uses to identify direct matches between a question and a table.

%Based on this obervation, we also represent a table with relations among Columns
%(entities) and the table Title. This process includes two steps. First, a table
%is decomposed into many small relations called \textit{Table Relations} and we
%call the decomposition Table Linearization. Second,  each \textit{Table
%Relation} is encoded into a dense vector in semantic space and we call it Table
%Relation Encoding.    

\mypar{Row-Wise Complete Graph (\textsf{RCG})} 
%\update{R2O4 \\ R2A2}{
Representing a table as triples is common practice in the semantic web \mbox{\cite{allemang2020semantic}} community. One challenge of doing so is to identify correctly the entity and relationships the table represents. With an entity-relationship diagram available (assumed known for semantic web), this is straightforward, but we do not have access to those in discovery scenarios with large collections of tables. Instead, we use a complete graph to connect each pair of cells in each row in the table,
%} 
as shown in the 
\F\ref{fig:table_repre_example}. Furthermore, the table title is included as a
special column of each table, thus it is also represented in triples for every
row. In such complete graph, some edges will be spurious, while others will
match the table question, as shown in the example. We are not concerned with
the spurious relationships because the first-stage retrieval will filter some out
if they do not match any question, and the relevance model in Section~\ref{subsec:relevance_model} will learn to distinguish more. 
We show empirically in the evaluation the benefits of such a representation. Next we explain how to encode the graph for
first-stage retrieval.

%If the entity relations in a table are given, e.g.
%a ER diagram describing the relationship between two columns, this decomposition
%would be much easier. However, in table discovery, a table may come from
%different sources and such relation information may be not available. So we
%propose a generic approach and assume no such meta data so that our system can
%be applied to  more scenarios. 
%Specifically, we organize the cells for each row
%and the table Title into a complete graph which include as many true relations
%implied in a table as possbile. In the graph each table cell is a node and the
%table Title is a special node.  An example of graph and table is shown in \F
%\ref{fig:table_repre_example}. We then treat each edge together with the two end
%nodes a \textit{Table Relation} which descrbes the relation between these two
%nodes. Since this is a complete graph, some \textit{Table Relations} may
%represents a true relation implied in the table , while many others are just
%noise, e.g., \textit { (Mon. \& Wed., 10-6; ...,  Hours of Operation - City,
%Chicago)}. This noise is often not a issue, because the true relations
%contributes more to the question. We call our table linearization strategy
%row-wise complete graph (RCG). 

\mypar{Encoding} In this last step, the goal is to encode each triple (edge) from the
graph produced by \textsf{RCG}. First we convert each triple into a snippet of text. 
Specifically,  a triple text always begins with the table title and is then followed by column name and cell value of \texttt{subject} and \texttt{object}.
Double quotes and commas are added to indicate the boundary. 
E.g. The triple text for \textit{ (Employment Resource Center , Facility - Hours , M-F: 9AM-5PM; SA: 9AM-1PM)} is 
\textit{"Public Technology Resources" ,  Facility  "Employment Resource Center"  ,  Hours  "M-F: 9AM-5PM; SA: 9AM-1PM".}
Second, to encode the tripe text, we use dense representations because we
found they outperform sparse representations based on \textsf{TF-IDF} and
\textsf{BM25}, as we show in the evaluation section.  We use vector
representations provided by a student dense retrieval model distilled from a pre-trained teacher dense retrieval over text (\cite{izacard2020distilling}), 
we leave why and how to distill the student model to Section~\ref{sec:implementation} . 
The student model contains a question encoder
(which we use to encode questions) and a passage encoder to encode the triples.
The resulting dense vectors for triples are then indexed for first-stage retrieval.

\subsection{Relevance Model}
\label{subsec:relevance_model}

We design a relevance model to solve the second-stage ranking problem, thus
addressing \textbf{Challenge 3 (C3)}. Given a question, $q$, the first-stage
retrieval returns $K_u$ triples arising from $K_t$ tables (there may be multiple
triples from the same table). The goal of second-stage ranking is to choose one
table $T_i$ from $K_t$. The general idea is to match $q$ to
triples from $T_i$. To do so, we use a pretrained \textsf{QA} model over text
\cite{izacard2020distilling} to match $q$ to each triple, and then construct a
matching representation ($q$ ,$T_i$). The \textsf{QA} model takes a question
and a text representation of a triple and outputs a matching representation.
We employ two techniques to boost the matching: \textbf{triple annotation} and
\textbf{representation augmentation}. 

%To enhance the representation, we annotate
%the triple (\textbf{Triple Annotation} below). Second, multiple augmented
%$(q,T_i)$ representations are constructed from the base  $(q,T_i)$
%representation and each triple to alleviate the distribution shit between
%synthetic and test data. Then each augmented $(q,T_i)$ is scored and the
%highest is picked as the choice. In the following, we will first introduce
%triple annotation, and then representation augmentation and finally the model
%architecture.

\mypar{Triple annotation} We use the pretrained QA model
over text \cite{izacard2020distilling} to provide input features for each (question,
triple) pair. To use the \textsf{QA} model, we convert triples to
text, but this transformation to text is different than the performed during
first-stage retrieval. 
%\update{R3O3}{
Here, we seek to annotate the triple to improve its
context which, in turn, helps with ranking as such context often includes sufficient information for the model to resolve semantic differences between words. To improve the context, we use
special tags to indicate the role of each string element in the triple and in
the table. For example, 
consider the aggregation question: \textit{"Which public technology resource in Chicago has the
longest hours of operation?''}. A triple in \mbox{\F\ref{fig:table_repre_example}}
that refers to the column \textit{``Hours''} is more likely to
answer the question than a triple from another table where the text
\emph{``Hours''} appears in a cell. We incorporate the context as
part of the input fed to the \textsf{QA} model: \textsf{[\_T\_]} to denote the
table Title; \textsf{[\_SC\_]} to denote subject column name $c_x$. \textsf{[\_S\_]} for
$Cell(r, c_x)$, which means Subject. \textsf{[\_OC\_]} for column name $c_y$, which
indicates Object Column. \textsf{[\_O\_]} for $Cell(r, c_y)$, which indicates Object.
%}

When a triple involves two columns, we still include the table title as context.
For example, the triple \textit {(Employment Resource Center , Facility - Hours , M-F: 9AM-5PM; SA: 9AM-1PM))} 
is annotated as ``[\_T\_] Public Technology Resources [\_SC\_] Facility [\_S\_]
Employment Resource Center [\_OC\_] Hours [\_O\_] M-F: 9AM-5PM; SA: 9AM-1PM''. In contrast, the
triple \textit{(Public Technology Resources, Facility, Employment Resource Center)} that only refers
to one column is annotated as: ``[\_T\_] Public Technology Resources [\_SC\_] [\_S\_] [\_OC\_] Facility
[\_O\_] Employment Resource Center''. That is, ``Public Technology Resources'' works as
title and subject. 

\mypar{Representation augmentation} We augment the representation by adding
redundancy~\cite{tobin2017domain}, which is shown to reduce distributional shift
and improve the matching process. A table is represented by the triples
retrieved during the first-stage retrieval stage: $(q, \{p_1, \cdots, p_m\})$.
To augment that representation, we represent the pair $m$ times, including each
time the question and each of the triples, i.e., $(q, \{p_1, \cdots, p_m\}) +
(q, p_i) \forall i \in m$. This redundancy helps to introduce diversity because
they share the label.

% \notefb{R2D2. The textual descriptions of 4.2/4.3 lacked a little precision and could be expanded for clarity. Example of lack of clarity:
% ... I wasn't sure what this means, particularly the projection statement. In both these sections, more math or a more formal description of the ML model might actually make things clearer. The row-wise complete graph I also felt could use a bit more expansion.}

\mypar{Model} Given a question $q$ and $K$ annotated triples $p_1, p_2, \ldots
, p_{K} $, we first feed all of them to the pretrained \textsf{QA} model to
get a feature vector $X_i$ for each ($q$, $p_i$) pair, $X_i = qa\_enc(q, p_i)$.
% as shown in Equation~\ref{eq:x}. 
We do not change the parameters of \textsf{QA}, so $X_i$ is fixed. We then project $X_i$ to a vector space w.r.t. table to
get a vector $REP_t(q, p_i) = W_t X_i + b_t$ and then apply element-wise max-pooling to all $REP_t(q, p_i)$
vectors in the same table to get an $(q, T_j)$ representation vector $REP(q, T_j) $:
\begin{align*}
REP(q, T_j) = Max\-Pool([REP_t(q, p^{T_j}_1) ; REP_t(q, p^{T_j}_2) ...])
\end{align*}

where $T_j$ indicates the table $X_i$ belongs to. Max-pooling helps select the best suitable triple $p_i$ for the question. To construct multiple
relevance representation, each $X_i$ is projected to another vector space
w.r.t. triple to get a vector $REP_u(q, p_i) =W_u X_i + b_u$  which is concatenated with
$REP(q, T_j)$ to get multiple equivalent relevance representation $REP(q, p_i, T_j)$: 
\begin{align*}
REP(q, p_i, T_j)=[REP(q, T_j) ; REP_u(q, p_i)]
\end{align*}

Then a multilayer perceptron (MLP) \cite{Goodfellow-et-al-2016} layer is added to compute the relevance score $s(q, p_i,)=MLP(REP(q, p_i,T_j))$ and then a Sigmoid layer and Binary Cross Entropy is added to it to compute 
the maximum likelihood estimation $\log {P \left (D|\theta \right )}$:
\begin{align*}
\log {P \left (D|\theta \right )} = \sum_{i=1}^{m}(y_i\log\sigma(s(q, p_i)) + (1\text{-}y_i)\log(1\text{-}\sigma(s(q, p_i))))
\end{align*}

where $y_i$ is the the 0/1 label. 
%\update{R3O3}{
During prediction, tables are ranked according to a relevance score computed as $\max_{p_i \in T_j} REP(q, p_i, T_j)$. The larger the score the more relevant the table is to the query and thus the higher in the ranking the table will appear.
%}

To make max-pooling more effective, the learned table-perspective
representation $REP_t(q, p_i)$ must be as different from each other as possible in a table because each $p_i$ is supposed to be a different triple in that table. To this end, we add a regularizer, the Diversity Regularizer to the final loss to minimizes 
the mean of mutual dot product of  $REP_t(q, p_i)$ in a table.

\mypar{Summary of relevance type}
%\update{R3O3}{
Our relevance model ranks question/table relevance by using the feature vector encoded by the pretrained QA model \mbox{\cite{izacard2020distilling}} and these vectors are good representation of word semantics. In addition we incorporate table context (see the ``Triple Annotation'' paragraph), that results in the relevance model ranking higher those tables that are semantically more related (according to the context) to the question. The output ranking reflects the relevance of tables with the question. When questions are ambiguous and the context is not sufficient to disambiguate the meaning of the question, those relevant tables make it into the ranking, letting users resolve the remaining ambiguity. To be robust to both context-rich and context-poor scenarios, our question generation strategy (see Section~\mbox{\ref{subsec:syt_data_generation}}) samples different number of columns from a table and includes the title probabilistically, thus producing training datasets with different contexts. More columns with the title means more specific questions (more context), while fewer columns and no title means more ambiguous questions.
%}

%\begin{figure}[t] 
%\centering
%\includegraphics[width=\columnwidth]{img/relevance_model}
%\caption{Model of relevance between a question $q$ and triples from the same
%table returned by first-stage retrieval.  Triples $p_i, \cdots, p_K)$ come from
%two tables $T_j$ and $T_k$ . The function qa\_enc is the encoder in the
%\textsf{QA} model} 
%\label{fig:post_stage_model} 
%\end{figure}

\section{\sys Overview and Implementation}
\label{sec:implementation}

In this section, we first give an overview of our end-to-end system \sys to 
show how the components work together to solve the table discovery problem. 
The system is built in Python using around 5,000 lines of new code. 
The system contains a \textsf{CLI} and a \textsf{UI} interface for users to type questions and then show top $N$ tables in order of relevance to the question. Each result in the ranking is accompanied by the provenance of that table including what triples match the question, so users have an easier time understanding the results.

While building \sys to implement the self-supervised approach explained in the previous sections, we run into severe challenges that would render a learned discovery system impractical. Here, we explain those challenges and the optimizations we implemented to address them (Section~\ref{subsec:techchallenges}).
%followed by other important technical details in Section~\ref{subsec:others}. 
We start with an overview of \sys.
% Then we discuss the key challenges we found in building the system and how we address them followed by other implementation details.

% \notera{say how it's built, say number of lines of code, etc.}

\subsection{Overview}
\begin{figure}[t] 
\centering
\includegraphics[width=0.65\columnwidth]{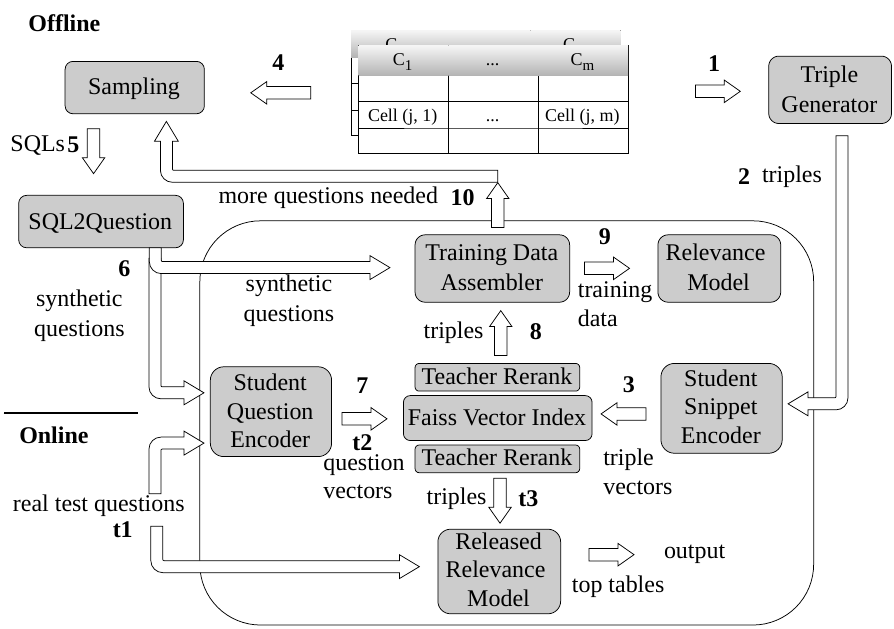}
\caption{Overview of \sys}
\label{fig:system_overview} 
\end{figure}

\F \ref{fig:system_overview} shows the overview of \sys,
which has an offline mode and online mode. In offline mode, tables are
decomposed into triples (step 1) which are then encoded (step 2), and indexed
(step 3). Then SQL queries are sampled from the target table collection (step 4) and
then a \textsf{SQL2Question} module translates SQL queries (step 5) into synthetic
questions which are encoded (step 6) and sent to a vector index (step 7, we use \textsf{Faiss} \cite{johnson2019billion} as it is scalable and supports similarity search). A small set of top triples are returned (step 8) for each synthetic question to
train the relevance model (step 9). If the more training data is needed, 
the training data assembler sends a message (step 10) for more questions.
In online mode, a real question is encoded (step t1) and sent  to
the vector index (t2). A small set of triples are returned and together with the
question are sent to the released relevance model (t3) to predict the most relevant
tables.

% \notera{explain the key challenges we found when we built the system. then, explain that you address those next, and give the outline for the next named paragraphs}

\subsection{Key challenges and solutions}
\label{subsec:techchallenges}
%\mypar{\textcolor {orange} {Question and triple encoder}} \notedel{ We use \textsf{FiD} retrieval
%\cite{izacard2020distilling}, a pretrained passage retrieval for question
%answering over question and free text, to encode question and triples into
%728-dimensional vectors. We use
%the version pre-trained on \textsf{TriviaQA} \cite{joshi2017triviaqa}, a
%high-quality dataset with 95K question answer pairs with evidence documents.
%}

A dense first-stage text retrieval encodes questions and text snippets (triple text in our case) separately 
so that the large amount of text snippets in a corpus can be pre-encoded before query. 
However, existing off-the-shelf pre-trained dense retrievals are often very expensive 
because of the transformer architecture \cite {vaswani2017attention} with many layers and so encoding can take very long time 
given a large amount of  triples. As an example, in the City of Chicago Open Data Portal, although the number of tables is not large, 
some tables have as many as 10 millions of rows and in total will result in around 42 billion triples. 
We experiment with 20 million triples on a single NVIDIA Quadro RTX 6000 GPU node and use the state-of-the-art  \textsf{FiD} 
dense retrieval \cite{izacard2020distilling} the version pre-trained on \textsf{TriviaQA} \cite{joshi2017triviaqa}, a
high-quality dataset with 95K question answer pairs with evidence documents. It takes around 6 hours to finish the encoding. 
To encode all 42 billion snippets of this dataset would take a prohibitive amount of 525 days. Using cloud GPUs to parallelize encoding is an expensive option.
 For example, at the time of this paper, Lambda \cite{lambdalabs} offers the Quadro RTX 6000 instance with \$0.5/h,
and to finish the encoding of 42 billion snippets in one day with
525 GPUs (often you are not allowed to launch so many instances) would cost 525 x 24 x 0.5  = \$6,300.  
Then either the user spends more money on machines or waits more time for encoding.  
\textit{So the first challenge is to implement a light encoder without affecting the system accuracy.} 

Another major challenge with dense vector representation is the storage footprint of the resulting vectors. Often each triple is encoded into a high dimensional vector of 768 floating point numbers. If each number is stored with 16-bit float precision, a vector will use  $768 \times (16 / 8) / 1024 = 1.5 $ k storage and 42 billion vectors will take up around 60T. This is clearly beyond the capacity of many organizations that would otherwise benefit from learned discovery systems. \textit{The second challenge is to reduce storage size without affecting the system accuracy}. Now we describe the solutions.
 
 \subsubsection{Light dense retrieval.} It contains two components:
 
\mypar{Student encoder} 
To get a cheap encoder, we use knowledge distillation \cite{hinton2015distilling} which uses the heavy Fid encoder (the teacher) to guide the training of a light student encoder. So the student encoder is supervised both from the \textsf{TriviaQA} data label (used for training the teacher) and the scores from the teacher on \textsf{TriviaQA}. In Fid, there is only one encoder for both question and passage. To minimize the accuracy drop during distillation, we use two with the student question encoder the same model architecture as Fid but the student text snippet encoder having only 1 layer versus 12 layers in Fid. With this design,  we get around 5x speedup in encoding. Another bottleneck of encoding is tokenization (on CPU) of input text which uses time close to the student model inference (on GPU) time. We then use multi-threading and the result student encoder is around 10x faster than the teacher also with multi-threading.
 
\mypar{Teacher to rerank} To compensate for the accuracy drop by the cheap student encoder, we use the teacher to rerank the n snippets from the student and output the top k ones to the relevance model.
 
 \subsubsection{Reduce storage size }
To reduce the storage, we use product quantization \cite{jegou2010product} where each vector is split into m (divisible by 768) parts and each part is an identifier (1 byte) of one of 256 clusters. The compression ratio is then (768*2+8)/(m+8) $\approx$ 768*2/m where the 8 bytes is the integer triple ID. To achieve comparable retrieval accuracy. E.g. with m=384, the student needs to retrieve n=1,500 triples for the teacher to rerank and 3,000 for m=256. More triples with smaller m will make the teacher rerank more expensive and also accuracy drops quickly. So m=384 is used by default, thus using 1/4 of the storage footprint.

\section{Evaluation}
\label{sec:evaluation}

% \notefb{R3O3. The evaluation is not comprehensive. The authors did not justify the choice of evaluation metrics of precision@K. Why not consider other metrics like recall@K, Hit@K or overall F1 score? Besides, since this work aimed at the problem of data discovery, a case study should also be provided to show some examples about the results discovered by the proposed framework.}

% \notefb{R3O4. More baseline methods needed to be compared with. One contribution claimed by the author is to employ self-supervised learning that avoids the requirement of labeled training data. There are many other design choices that can reach this goal. For example, use syntactic similarity and word embeddings as [28] did. There are also some pre-trained language models for tabular data and texts, such as:
% - TABERT: Pretraining for Joint Understanding of Textual and Tabular Data. ACL 2020.
% - TAPEX: Table Pre-training via Learning a Neural SQL Executor. ICLR 2022.
% It is easy to extend these methods as the baseline of the problem studied here: the text and column/row/cells of the table can be encoded into vectors by such pre-trained models. And the results can be found by calculating the cosine similarity between the embedding vectors.}

In this section, we answer the main research questions of our work:

% \begin{myitemize}

\noindent$\bullet$~\textbf{RQ1. Does self-supervised data discovery work?} The main claim of
our work is that learned discovery systems need repository-specific training
data to perform well, and that our self-supervised approach can collect that
data automatically. We measure the extent to which systems suffer when not
specifically trained for a target repository and we measure the performance of
our approach. We will compare with strong baselines to contextualize the
results. This is presented in Section~\ref{sec:rq1}.

\noindent$\bullet$~\textbf{RQ2. Does Bayesian incremental training work?} A challenge of
producing the training dataset is to choose its size. Too large will lead to
unnecessary resource consumption without any accuracy benefits and possibly a
degradation. Too small will lead to underperforming models. The technique we
introduce uses Bayesian neural networks to know when to stop. We evaluate its
effectiveness here when compared to a baseline approach that retrains
iteratively using ever larger datasets. This is presented in
Section~\ref{sec:rq2}.

\noindent$\bullet$~\textbf{RQ3. How does table representation affect accuracy?} An important
contributor to the end to end performance is the representation of tables. We
argued in the introduction that many existing approaches do not work well and
thus we introduced a new row-wise graph based approach. See Section~\ref{sec:rq3}.

\noindent$\bullet$~\textbf{RQ4. Microbenchmarks} To provide a full account of the performance
of the end to end system we include microbenchmarks that concentrate in
answering two questions: the performance differences of using sparse vs dense
indices during the first-stage retrieval, and the runtime performance of the new
approach, accounting for both offline and online components. See Section~\ref{sec:rq4}.

% \end{myitemize}

\mypar{Metrics} The ultimate goal is to identify the table that
contains the answer to an input question. To measure the accuracy of the
different baselines when given a set of questions we use the precision-at-K
metric (P@K) aggregated over all questions, as in previous
work~\cite{herzig2021open}. This metric indicates the ratio of questions for
which the answer is in the top K tables. We measure P@1 and P@5. 
We do not use recall@K because in the benchmarks we use, each question has 1 single table as the answer. More specifically, in one benchmark there is exactly one table per question, while in the other only 7 questions out of 959 have 2 valid answer tables. Thus, measuring precision alone is sufficient to understand the quality of the approaches we compare.

\mypar{Datasets} We use two benchmark datasets that have been extensively used
by previous systems and that have been generated from real-world representative
queries. 

% \begin{myitemize}
%\noindent$\bullet$ 
\noindent$\bullet$~\textbf{NQ-Tables}~\cite{herzig2021open} This dataset
contains 210,454 tables extracted from Wikipedia pages and 959 test questions
which are a subset of the Google Natural Questions \cite{google_nq}. It is very
popular in table retrieval \cite{shraga2020web, pan2021cltr, herzig2021open}
because the questions are asked by real Google Search users. The tables are
relatively dirty, with 23.3\% tables having some column header missing and 63\%
tables having cells that contain long chunks of text.
%with much longer text by IQR
%outlier detection \cite{peter2021srm}  which make retrieval much more
%challenging. 

%\noindent$\bullet$ 
\noindent$\bullet$~\textbf{FetaQA}~\cite{nan2022fetaqa} This dataset contains
10,330 clean tables from Wikipedia and 2003 test questions.  The questions are
generally longer and more complex than those from \textsf{NQ-Tables}.

% \end{myitemize}

\mypar{System Setup} We run all experiments on the Chameleon Cloud \cite
{keahey2020lessons}. We use one node with 48 processing threads, 187G RAM, and
an NVIDIA RTX 6000 GPU. The OS is Ubuntu 18.04 and the CUDA version is 11.2 and
10.2 (for other baselines). The system is implemented in Python v3.7.9 and
uses pytorch 1.12.1.

\subsection{RQ1. Self-supervised data discovery}
\label{sec:rq1}

We measure the P@1 and P@5 accuracy of the self-supervised discovery system (
%\update{R2O2}{
that uses the synthetically generated training data
%}
) on the two benchmark datasets. To interpret the results, we compare them against
the following baselines:

%In this section, we evaluate the performance of our end-to-end self-supervised
%system agaist other state-of-the-art systems trained on human-annonated data,
%synthetic data and also system that does not need to be trained.  We first
%introduce the baselines, and then present the results.

%\mypar{Baselines} We measure the retrieval accuracy of our system against the
%following baselines.  

\mypar{BM25 (Table-Row-Tokens)} Discovery
systems such as Aurum~\cite{fernandez2018aurum} and
Auctus~\cite{castelo2021auctus} use traditional information retrieval techniques
based on BM25 to retrieve tables given keywords. We implement this baseline to
represent these discovery solutions. In particular, we index every row of a table, along with the table's title in Elastic~\cite{elasticsearch}.

\mypar{Tabert} \cite{yin2020tabert} is pretrained on 26 million tables 
with corresponding free text context. It does not solve data discovery, instead it answers questions by pointing to the cell of a given input table. Still, we use this model to understand if its table representation works for data discovery. We emphasize that Tabert encodes a question and table pair simultaneously and thus does not scale to large repositories of tables. We use \emph{our} first-stage retrieval system to get a small set of tables as candidates. And then we use Tabert to output a question embedding and a column embedding for each column in a table. We use the mean column embedding and compute the cosine similarity score of the question and mean column embedding to rank tables.

\mypar{OpenDTR}~\cite{herzig2021open} is a state-of-the-art
learned discovery system.  We use 3 variants: \textsf{OpenDTR-Retrain},
\textsf{OpenDTR-NoRetrain} and \textsf{OpenDTR-Synthetic}.

% \begin{myitemize}

\noindent$\bullet$~\textsf{OpenDTR-Retrain} is retrained on a human-annotated training dataset that
has the same distribution as the test dataset. For example, when the target test
dataset is \textsf{NQ-Tables}, \textsf{OpenDTR-Retrain} is trained on
\textsf{NQ-Tables}. This \emph{baseline requires collecting
training data for each dataset and is thus not desirable}. Still, we include it
because it allows us to understand the performance difference between
(expensive) human-collected training datasets and our approach.

\noindent$\bullet$~\textsf{OpenDTR-NoRetrain} is the system instance pretrained on a
human-annotated training dataset having different distribution from the test
dataset. For example, when the test dataset is \textsf{NQ-Tables},
\textsf{OpenDTR-NoRetrain} means the system instance trained on \textsf{FetaQA}.
This baseline helps us understand what happens to the system performance when a
learned discovery system is deployed on a new table collection without retraining.
% Our claim is that the performance will degrade and that, in turn,
% justifies the self-supervised approach that: i) yields good performance while;
% ii) avoiding human cost of collecting training data.

\noindent$\bullet$~Finally, \textsf{OpenDTR-Synthetic}
is trained on the synthetic questions generated by our self-supervised approach. We use this baseline to understand the impact of the new table representation and the semantic relevance model.

% \end{myitemize}

\mypar{GTR}~\cite{wang2021retrieving} is a state-of-the-art
second-stage ranking model. Because it only solves second-stage ranking, we use
our first-stage retrieval system to obtain results end-to-end. We implement
3 variants as before: \textsf{GTR-Retrain}, \textsf{GTR-NoRetrain} and
\textsf{GTR-Synthetic}.

\mypar{Experimental Setup} 
%\textcolor{blue} { 
%Since both \textsf{OpenDTR} and \textsf{GTR} use maximum likelihood estimation (MLE)
%for training, i.e. $\tilde{\theta} = \argmax {P(D|\theta)}$. 
%We also trains the MLE-version relevance model using fixed dataset and call the whole self-supervised system
%\textsf{SL2D-MLE}. Sepcifically, we use a synthetic dataset of size 10,000 (questions), same order of real training data size provided in the two datasets 
%\textsf{NQ-Tables} and \textsf{FetaQA}. This synthetic dataset are also used by 
%\textsf{BM25 (Table-Row-Tokens)}, \textsf{OpenDTR-Synthetic} and \textsf{GTR-Synthetic}. 
%}
We train OpenDTR models using the released official
code. In our system, we encode each triple as a 768-dimensional vector using the
Fid Retriever~\cite{izacard2020distilling} pretrained on the TriviaQA dataset.
This results in 41,883,973 vectors on NQ-Tables, and 3,877,603 vectors on
FetaQA. We index those vectors using Faiss \cite{johnson2019billion}. We
retrieve at least 5 tables during first-stage retrieval and then apply the
second-stage ranking. We use the original code to train the \textsf{GTR} models. We train
our system, \textsf{OpenDTR-Synthetic} and \textsf{GTR-Synthetic} using the exact same set of
training data produced in a self-supervised manner.

\begin{figure}[ht]
    \includegraphics[width=0.65\columnwidth]{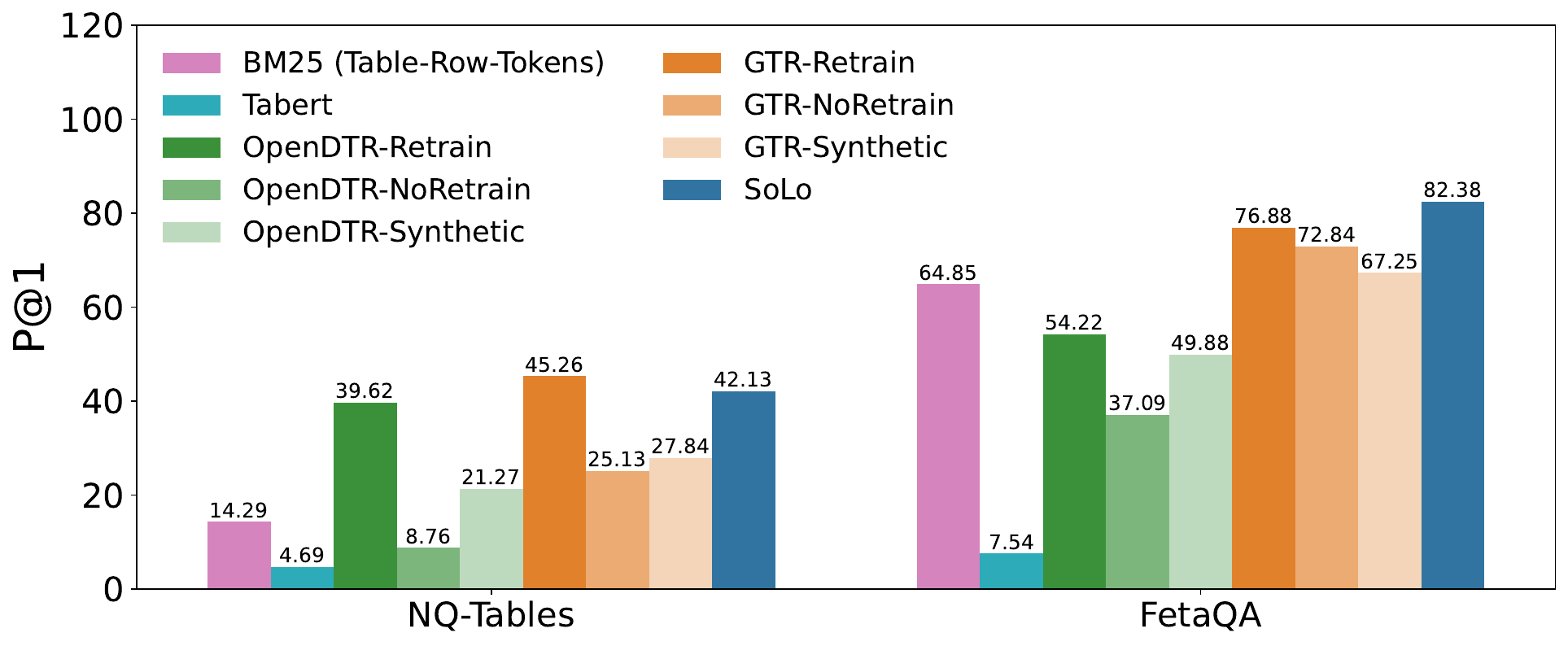}
    \caption{P@1 Accuracy on baseline datasets}
	\label{fig:system_accuracyp1}
\end{figure}

\begin{figure}[ht]
    \centering
    \includegraphics[width=0.65\columnwidth]{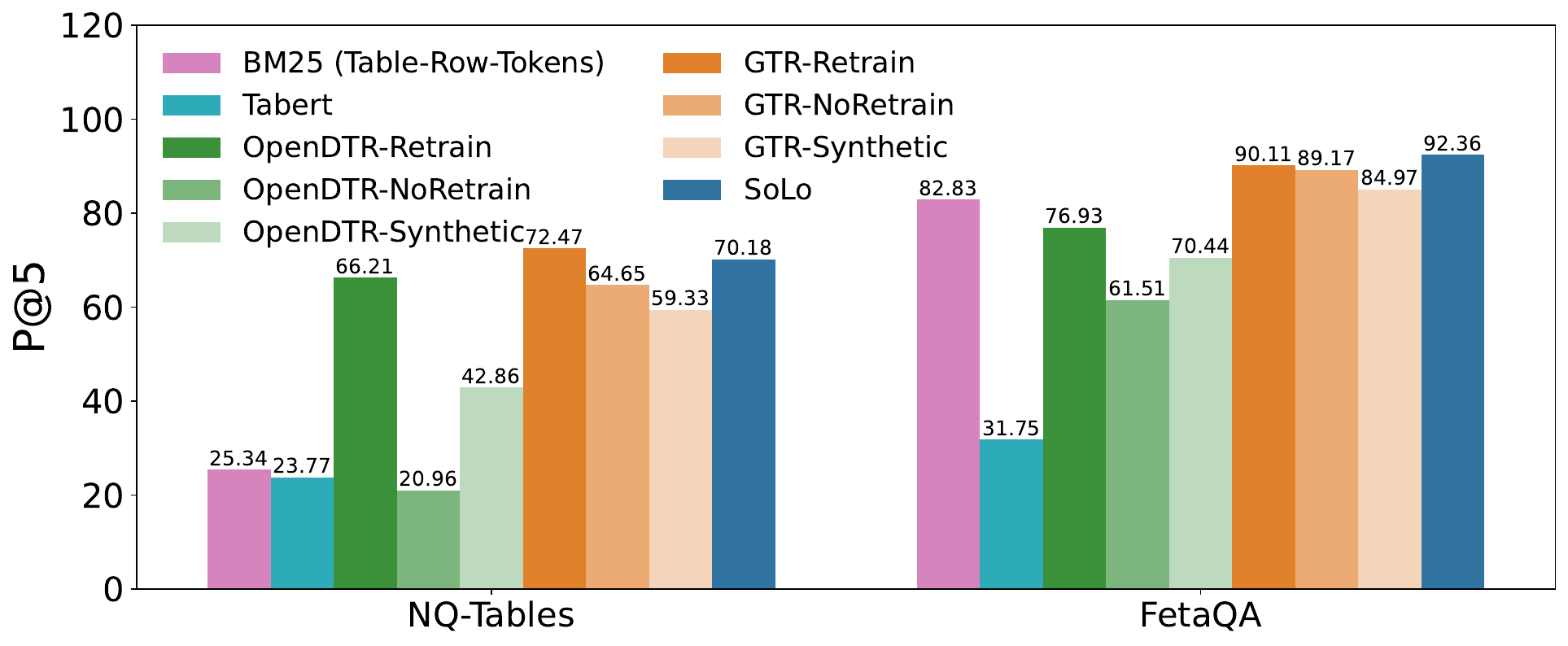}
    \caption{P@5 Accuracy on baseline datasets}
	\label{fig:system_accuracyp5}
\end{figure}

\subsubsection{Main Results}

We show the P@1 and P@5 in \F\ref{fig:system_accuracyp1} and
\F\ref{fig:system_accuracyp5}, respectively. We highlight several key insights:

\mypar{Baselines underperform on new table collections when not specifically
retrained} The first observation we make concerns \textsf{OpenDTR-NoRetrain} and
\textsf{GTR-NoRetrain}. When these systems are trained with data from one
benchmark and evaluated on the other, their performance deteriorates
significantly, by 30 and 18 points in the case of \textsf{OpenDTR} on
\textsf{NQ-Tables} and \textsf{FetaQA}, respectively. And by 20 and 4 points in
the case of \textsf{GTR-NoRetrain}. This demonstrates that without retraining, a
learned table discovery system vastly underperforms on new table collections.

\mypar{Off-the-shelf pretrained Table QA representation does not generalize to table discovery}
The very low accuracies of \textsf{Tabert} show pretrained embeddings in table scenario are hard to applied to related tasks and training is needed to get good performance.

\mypar{\sysmle produces synthetic training datasets that achieve high accuracy}
\sysmle vastly outperforms non-trained systems. Compared to the
\textsf{NoRetrain} variants, \sysmle achieves 30 and 15 more points than
\textsf{OpenDTR-NoRetrain} and \textsf{GTR-NoRetrain} in \textsf{NQ-Tables} and
40 and 13 more points in \textsf{FetaQA}. This improvement in accuracy comes at
the same cost for the human, who does not need to collect data and label it
manually because our approach does it for them. This result validates the main
contribution of our paper.

%\mypar{test}\sysmle outperforms any system that is not specifically
%retrained on the target dataset (\textsf{OpenDTR-NoRetrain} and
%\textsf{GTR-NoRetrain}). The performance gap ranges from 45 to 10 points
%across datasets and systems in P@1 and correspondingly larger in P@5. The
%self-supervised approach achieves much higher accuracy than the other baselines
%without human intervention, validating the main contribution of
%the paper that addresses Challenge 1.

\mypar{\sysmle is competitive in performance compared to the retrain
baselines without needing to pay the cost of obtaining a new training dataset}
Even when compared to the other baselines retrained on benchmark-specific
training datasets, \sysmle achieves good performance. In fact \sysmle
outperforms \textsf{OpenDTR-Retrain} in both datasets by 3 and 25 points in
\textsf{NQ-Tables} and \textsf{FetaQA} respectively. It also outperforms
\textsf{GTR-Retrain} in the \textsf{FetaQA} benchmark by 6 points. It
underperforms \textsf{GTR-Retrain} in the \textsf{NQ-Tables} dataset by only 3
points, but again, to emphasize, without paying the cost of collecting data
manually.

\mypar{\sysmle always outperforms \textsf{BM25}, but this is not true for the
other baselines} \textsf{BM25} represents the retrieval performance of
non-learned discovery systems. Note that \textsf{OpenDTR-NoRetrain}
underperforms \textsf{BM25} in \textsf{NQ-Tables} by 6 points. This means that,
without our approach, and without the ability to collect a new dataset, a
non-learned data discovery solution will outperform the more sophisticated
\textsf{OpenDTR}. Or rather, that collecting high-quality training data is
decidedly important for learned table discovery to perform well on new table
collections. \textsf{GTR} does better than \textsf{OpenDTR} when compared to
\textsf{BM25}---recall it uses our new first-stage retrieval. In contrast,
\sysmle always outperforms \textsf{BM25}.

\mypar{\sysmle performance benefits go beyond the synthetic data generation}
To test this hypothesis and evaluate the contributions of the new table
representation and relevance model we use the same synthetically generated
dataset produced by our approach to train all baselines and compare their
performance; this corresponds to the \textsf{Synthetic} baselines. As shown in
the figures, when all baselines are trained on the synthetically generated
dataset, \sysmle outperforms every other baseline by 20, 15, 33, and 15 points
(from the left to the right of the figure). These results validate the design of
the table representation and retrieval model.

\mypar{The trends are similar and accentuated when measuring P@5} The trend when
observing P@5 is the same as in P@1, with the total accuracy much higher, as
expected. This suggests that when the end user has the bandwidth to manually
check a ranking of 5 tables that may answer their question, they are much more
likely to identify the answer they are after.

\subsubsection{In-depth analysis of results} We ask: 
\mypar{Why do the \textsf{Retrain} baselines perform worse than \sysmle despite
having access to high-quality manually collected training data?} When analyzing
the logs for the \textsf{FetaQA} benchmark, we find that \sysmle performs better
at matching entities in the question and table at the word level. This is
largely because \sysmle takes advantage of the OpenQA
\cite{izacard2020distilling} model which is pretrained on the TriviaQA
\cite{joshi2017triviaqa} question dataset. The same reason applies to
\textsf{OpenDTR-Retrain}. 
%\update{R2O5}{
Note, it is common practice nowadays to transfer knowledge from a model pretrained on high-quality human data \mbox{\cite{qiu2020pre}}; but this is orthogonal to {\sys}'s users, who do not need to annotated data manually.
%}

On \textsf{NQ-Tables} (where \sysmle underperforms \textsf{GTR-Retrain}), we
find \sysmle is more likely misled by long cells that contain information related 
to the question, while
\textsf{GTR-Retrain} performs better in
exploiting the table cell topology structure to find the ground truth table. We
use an indicative example from our logs. Given the question \textit{``where is
hindu kush mountains located on a map''}, \sysmle chooses a table with a long
cell text \textit{``The general location of the Himalayas mountain range (this
map has the Hindu Kush in the Himalaya, not normally regarded as part of the
core Himalayas)''}. The text does indeed contain an answer. In contrast,
\textsf{GTR} chooses a table with
a cell containing \textit{``Hindu Kush''} and a cell containing the coordinates,
which gives a more precise answer and is what the benchmark was expecting. It is
arguable whether one answer is indeed superior to the other for an end-user, but
it is certainly the case the benchmark favors \textsf{GTR} in this case.
Finally, because \textsf{FetaQA} is a relatively clean benchmark and most cells
contain simple information, we do not see the same effect in this dataset, i.e.,
\sysmle is less likely to select a long cell and it consequently outperforms 
\textsf{GTR-Retrain} and \textsf{OpenDTR-Retrain}. Note that in any case, \emph{the
performance of \sysmle, which is close to the other baselines, is achieved
without any human-collected dataset.}

%\noindent$\bullet$ \sysmle outperforms the other baselines when trained on
%synthetic data (\textsf{OpenDTR-Synthetic} and \textsf{GTR-Synthetic}). In this
%case, the gap ranges from 30 to 20 points (P@1), thus, indicating that the table
%representation and relevance model are necessary to achieve high accuracy on a
%learned discovery system. This validates the technical solutions presented to
%address Challenges 2 and 3. 

\mypar{Why do the \textsf{Synthetic} baselines underperform?}
%\textcolor{blue} { \textsf{GTR-Retrain} can outperform \sysmle on
%\textsf{NQ-Tables} because the training data and test data come from the same
%distribution when using manual training data. Specifically, both the training
%data and test data are random samples from a manual dataset.  However,
\textsf{GTR} models tables as a graph where each cell is a node and it is
connected only to neighboring cells. This representation is unnatural for
relational data, where the order of columns does not matter. Such representation
makes the model brittle to situations where the distribution of the training
dataset and the test set differs, such as it is the case when comparing the
synthetically generated training dataset produced by our approach and the test
set of the benchmarks. Concretely, it is often the case that the subject and
object in a relation are not neighbors and thus do not make it into the
representation used by \textsf{GTR}. From our analysis we observe that
\textsf{GTR} is more likely to overfit on the synthetic data, thus explaining
the deterioration of quality. In contrast, our table representation does not
suffer from those problems, thus preventing overfitting. A similar phenomenon explains the results for
\textsf{OpenDTR-Synthetic}. Although this baseline does not use the same table
representation as \textsf{GTR}, it encodes question and table \emph{together},
using a pretrained TableQA~\cite{herzig2020tapas} model which seems brittle to the format of the training data.

%make it break when training data are
%synthetic and thus has different distribution from the manual test data.
%\textsf{GTR}  models a table as a graph where each cell is a node and connected
%only to neighboring cells in a table plus some virtual row nodes and column
%nodes . So often the subject and object in a relation are not neighboured but a
%few hops away. Given synthetic questions,  GTR learns the relevance by a graph
%neural network on tables induced by synthetic questions which definitly have
%different graph structure distribution from that of tables in test data. With a
%more complicated represenation model of tables, \textsf{GTR-Synthetic} more
%likely overfits on the synthetic training data. \sysmle instead uses a simpler
%representation of tables, i.e. triples which only includes subject, relationship
%and object.  A Simpler representaion than a graph will make \sysmle generlize
%better than \textsf{GTR-Synthetic} to unseen tables because the train/test
%triple difference definitrly has a much smaller upper bound than that of
%train/test graph difference.     }
%
%\textcolor{blue} { \textsf{OpenDTR-Synthetic} has the same issue. It takes the
%whole table structure as input to encode a question and table, and specifcally
%it uses a pretrained TableQA \cite{herzig2020tapas} model.  So the relevence
%representation is also more coupled with the table structure from synthetic
%questions  and performs worse on test data.  }

\subsection{RQ2. Bayesian Incremental Training}
\label{sec:rq2}

% \notefb{R2D4. The early stopping strategy is potentially over-aggressive it seems, or potentially some other wonkiness. It's very weird that the training is stopping reporting no improvement to validation data while test performance is substantially increasing.}

We measure the P@1(5) accuracy and training cost (the total
training time and epochs used) using the new Bayesian incremental training on the two
datasets and compare it against a baseline (\textsf{Simple}) that sequentially grows the
input training dataset and trains the model from scratch until it performs well.
% (as described in \ref{subsec:bnn_train}).

\mypar{Experimental Setup} We generate 10 partial training datasets $D_1,
\cdots, D_{10} $  for each benchmark, each $D_i$ with 
%\update{R2O6}{
1,000 different questions generated as in Section~\mbox{\ref{subsec:syt_data_generation}}
%}
with corresponding triples. Given a fixed $\{D_1, \cdots, D_m\}$, the simple
approach trains the relevance model (
%\update{R2A3}{
with state-of-the-art Xavier initialization of parameters as in Pytorch \mbox{\cite{paszke2017automatic}}
%}
) 
using maximum likelihood estimation and an
early-stop strategy with patience of 1 epoch (each epoch is a full pass of
$\{D_1, \cdots, D_m\}$). This means if a model checkpoint does not improve the
P@1 accuracy in two continuous epochs, the training process stops. The patience
of dataset for incremental training is also 1, i.e. if either $\{D_1, \cdots,
D_m,  D_{m+1}\}$ or $\{D_1, \cdots , D_m, D_{m+2}\}$ does not improve P@1 over
$\{D_1, \cdots, D_m\}$ the whole incremental training process stops.

The same patience of epoch and patience of dataset settings are applied to the
Bayesian approach. By ``Prior$(D1, \cdots , D_j), D_m$'' we indicate that $(D1,
\cdots , D_j)$ have been used for training previously and thus they constitute
the prior for training on $D_m$. During test, we sample 6 versions of model
parameters from the posterior distribution, and use also the posterior mean
parameters, then take the average of predictions of the 7 versions of parameters
as output.

% \usepackage{multirow}
\begin{table}[ht]
\begin{adjustbox}{width=0.65\columnwidth,center}
\begin{tabular}{|l|llrrrr|c|}
\hline
\multicolumn{1}{|c|}{\multirow{2}{*}{\textbf{Approach}}} & \multicolumn{1}{l|}{\multirow{2}{*}{\textbf{\begin{tabular}[c]{@{}l@{}}Time \\ Step\end{tabular}}}} & \multicolumn{1}{c|}{\multirow{2}{*}{\textbf{\begin{tabular}[c]{@{}c@{}}Training \\ Datasets\end{tabular}}}} & \multicolumn{2}{c|}{\textbf{Eval}}                                        & \multicolumn{2}{c|}{\textbf{Test}}                                      & \multirow{2}{*}{\textbf{\begin{tabular}[c]{@{}c@{}}Training Cost \\ S  (epochs)\end{tabular}}} \\ \cline{4-7}
\multicolumn{1}{|c|}{}                                   & \multicolumn{1}{l|}{}                                                                               & \multicolumn{1}{c|}{}                                                                                       & \multicolumn{1}{c|}{\textbf{P@1}}   & \multicolumn{1}{c|}{\textbf{P@5}}   & \multicolumn{1}{c|}{\textbf{P@1}}   & \multicolumn{1}{c|}{\textbf{P@5}} &                                                                                                \\ \hline
\multirow{7}{*}{Simple}                                  & \multicolumn{1}{l|}{t1}                                                                             & \multicolumn{1}{l|}{D1}                                                                                     & \multicolumn{1}{r|}{72.65}          & \multicolumn{1}{r|}{82.45}          & \multicolumn{1}{r|}{42.02}          & 70.80                             & 6,134 (7)                                                                                      \\ \cline{2-8} 
                                                         & \multicolumn{1}{l|}{t2}                                                                             & \multicolumn{1}{l|}{D1, D2}                                                                                 & \multicolumn{1}{r|}{73.70}          & \multicolumn{1}{r|}{82.75}          & \multicolumn{1}{r|}{42.65}          & 70.91                             & 9,258 (7)                                                                                      \\ \cline{2-8} 
                                                         & \multicolumn{1}{l|}{t3}                                                                             & \multicolumn{1}{l|}{D1, D2, D3}                                                                             & \multicolumn{1}{r|}{74.10}          & \multicolumn{1}{r|}{82.45}          & \multicolumn{1}{r|}{40.15}          & 70.59                             & 12,168 (7)                                                                                     \\ \cline{2-8} 
                                                         & \multicolumn{1}{l|}{t4}                                                                             & \multicolumn{1}{l|}{\textbf{D1, D2, D3, D4}}                                                                & \multicolumn{1}{r|}{\textbf{74.35}} & \multicolumn{1}{r|}{\textbf{83.05}} & \multicolumn{1}{r|}{\textbf{40.15}} & \textbf{69.76}                    & 21,650 (10)                                                                                    \\ \cline{2-8} 
                                                         & \multicolumn{1}{l|}{t5}                                                                             & \multicolumn{1}{l|}{D1, D2, D3, D4, D5}                                                                     & \multicolumn{1}{r|}{73.00}          & \multicolumn{1}{r|}{82.30}          & \multicolumn{1}{r|}{42.54}          & 70.91                             & 7,829 (3)                                                                                      \\ \cline{2-8} 
                                                         & \multicolumn{1}{l|}{t6}                                                                             & \multicolumn{1}{l|}{D1, D2, D3, D4, D6}                                                                     & \multicolumn{1}{r|}{72.65}          & \multicolumn{1}{r|}{82.80}          & \multicolumn{1}{r|}{42.96}          & 71.74                             & 7,752 (3)                                                                                      \\ \cline{2-8} 
                                                         & \multicolumn{6}{c|}{Sum}                                                                                                                                                                                                                                                                                                                                                & \textbf{64,790}                                                                                \\ \hline
\multirow{5}{*}{Bayesian}                                & \multicolumn{1}{l|}{t1}                                                                             & \multicolumn{1}{l|}{D1}                                                                                     & \multicolumn{1}{r|}{66.30}          & \multicolumn{1}{r|}{81.60}          & \multicolumn{1}{r|}{41.71}          & 70.70                             & 2,657 (3)                                                                                      \\ \cline{2-8} 
                                                         & \multicolumn{1}{l|}{t2}                                                                             & \multicolumn{1}{l|}{\textbf{Prior (D1), D2}}                                                                & \multicolumn{1}{r|}{\textbf{69.70}} & \multicolumn{1}{r|}{\textbf{82.35}} & \multicolumn{1}{r|}{\textbf{43.07}} & \textbf{71.32}                    & 7,221 (8)                                                                                      \\ \cline{2-8} 
                                                         & \multicolumn{1}{l|}{t3}                                                                             & \multicolumn{1}{l|}{Prior (D1, D2), D3}                                                                     & \multicolumn{1}{r|}{69.65}          & \multicolumn{1}{r|}{82.50}          & \multicolumn{1}{r|}{41.40}          & 71.32                             & 3,553 (4)                                                                                      \\ \cline{2-8} 
                                                         & \multicolumn{1}{l|}{t4}                                                                             & \multicolumn{1}{l|}{Prior (D1, D2), D4}                                                                     & \multicolumn{1}{r|}{69.40}          & \multicolumn{1}{r|}{82.25}          & \multicolumn{1}{r|}{40.88}          & 70.70                             & 2,679 (3)                                                                                      \\ \cline{2-8} 
                                                         & \multicolumn{6}{c|}{Sum}                                                                                                                                                                                                                                                                                                                                                & \textbf{16,108}                                                                                \\ \hline
\end{tabular}
\end{adjustbox}
\caption{Bayesian/Simple training results on NQ-Tables}
\label{tab:nq_tables_incremental_training}
\end{table}

% \notefb{R2D3 response. I meant if the naive solution was trained from scratch. I wasn't sure whether the starting parameters were taken from the end of the previous dataset training or from a normal NN initialization. Given how fast say t5,t6 trained in table 2 compared to t4, I figured some form of warm initialization was provided.}

% \notefb{R2D4 response. You should take your validation data from the same distribution as your test data. This makes more sense to me as its the performance on test data you care about.}
 
\mypar{Results} Table~\ref{tab:nq_tables_incremental_training} and
~\ref{tab:fetaqa_incremental_training} shows the results on \textsf{NQ-Tables}
and \textsf{FetaQA}. We summarize the insights:

\noindent$\bullet$ Our new Bayesian incremental approach takes much less training
time and achieves better test accuracy than the simple approach.
On \textsf{NQ-Tables}, the simple approach takes 64,790 seconds for the whole
incremental training and chooses $\{D1, D2, D3, D4\}$ on the evaluation
dataset. In contrast, the Bayesian incremental approach takes only 16,108
seconds (chooses $\{D1, D2\}$), a reduction of runtime of 4x, or equivalently 18 hours vs only 4 hours
to find a train the system for a new table collection. This reduction in runtime
makes it more feasible to keep the system up to date as the underlying data
tables naturally change. On \textsf{FetaQA}, the simple approach takes 25,028 seconds,
while the Bayesian incremental approach takes only 15,554 seconds, about 62\% of simple approach
with close P@1 (82.73 vs 82.53). 
%The main reason why simple approach cannot
%outperform Bayesian approach with approximation is that we are using a synthetic
%evaluation dataset which has a different distribution than the real test
%dataset. As in table ~\ref{tab:nq_tables_incremental_training}, on
%\textsf{NQ-Tables}, Simple approach achieves better  P@1 (74.35 vs 69.70) on the
%evaluation dataset, but perform worse on the test dataet, same for
%\textsf{FetaQA} as in table ~\ref{tab:fetaqa_incremental_training}. In addition,
%Bayesian approach performs better in generlization to unsee data because it
%introduce randomness explicitly in the model.

\noindent$\bullet$ More data does not necessarily lead to better
accuracy. Since the training data are generated automatically, there is 
redundancy and noise. As shown in
Table~\ref{tab:nq_tables_incremental_training}, $\{D1, D2, D3\}$ performs worse
than $\{D1, D2\}$ and even $\{D1\}$. Simply adding more data to the simple
approach does not suffice and also it slows down the entire process. This
further validates the Bayesian incremental approach.

\captionsetup{justification=centering}

% \usepackage{multirow}
\begin{table}[ht]
\begin{adjustbox}{width=0.65\columnwidth,center}
\begin{tabular}{|l|llcccc|c|}
\hline
\multicolumn{1}{|c|}{\multirow{2}{*}{\textbf{Approach}}} & \multicolumn{1}{l|}{\multirow{2}{*}{\textbf{\begin{tabular}[c]{@{}l@{}}Time\\ Step\end{tabular}}}} & \multicolumn{1}{c|}{\multirow{2}{*}{\textbf{\begin{tabular}[c]{@{}c@{}}Training \\ Datasets\end{tabular}}}} & \multicolumn{2}{c|}{\textbf{Eval}}                                        & \multicolumn{2}{c|}{\textbf{Test}}                   & \multirow{2}{*}{\textbf{\begin{tabular}[c]{@{}c@{}}Training Cost \\ S (epochs)\end{tabular}}} \\ \cline{4-7}
\multicolumn{1}{|c|}{}                                   & \multicolumn{1}{l|}{}                                                                              & \multicolumn{1}{c|}{}                                                                                       & \multicolumn{1}{c|}{\textbf{P@1}}   & \multicolumn{1}{c|}{\textbf{P@5}}   & \multicolumn{1}{c|}{\textbf{P@1}}   & \textbf{P@5}   &                                                                                               \\ \hline
\multirow{6}{*}{Simple}                                  & \multicolumn{1}{l|}{t1}                                                                            & \multicolumn{1}{l|}{D1}                                                                                     & \multicolumn{1}{c|}{76.70}          & \multicolumn{1}{c|}{80.25}          & \multicolumn{1}{c|}{82.88}          & 93.06          & 2,519 (4)                                                                                     \\ \cline{2-8} 
                                                         & \multicolumn{1}{l|}{t2}                                                                            & \multicolumn{1}{l|}{D1, D2}                                                                                 & \multicolumn{1}{c|}{77.05}          & \multicolumn{1}{c|}{80.20}          & \multicolumn{1}{c|}{83.08}          & 93.01          & 7,147 (8)                                                                                     \\ \cline{2-8} 
                                                         & \multicolumn{1}{l|}{t3}                                                                            & \multicolumn{1}{l|}{\textbf{D1, D2, D3}}                                                                    & \multicolumn{1}{c|}{\textbf{77.15}} & \multicolumn{1}{c|}{\textbf{80.25}} & \multicolumn{1}{c|}{\textbf{82.53}} & \textbf{93.01} & 6,889 (6)                                                                                     \\ \cline{2-8} 
                                                         & \multicolumn{1}{l|}{t4}                                                                            & \multicolumn{1}{l|}{D1, D2, D3, D4}                                                                         & \multicolumn{1}{c|}{76.35}          & \multicolumn{1}{c|}{79.90}          & \multicolumn{1}{c|}{82.63}          & 92.96          & 4,253 (3)                                                                                     \\ \cline{2-8} 
                                                         & \multicolumn{1}{l|}{t5}                                                                            & \multicolumn{1}{l|}{D1, D2, D3, D5}                                                                         & \multicolumn{1}{c|}{76.80}          & \multicolumn{1}{c|}{80.05}          & \multicolumn{1}{c|}{83.33}          & 93.06          & 4,220 (3)                                                                                     \\ \cline{2-8} 
                                                         & \multicolumn{6}{c|}{Sum}                                                                                                                                                                                                                                                                                                                            & \textbf{25,028}                                                                               \\ \hline
\multirow{5}{*}{Bayesian}                                & \multicolumn{1}{l|}{t1}                                                                            & \multicolumn{1}{l|}{D1}                                                                                     & \multicolumn{1}{c|}{75.85}          & \multicolumn{1}{c|}{80.00}          & \multicolumn{1}{c|}{82.43}          & 92.66          & 3,225 (5)                                                                                     \\ \cline{2-8} 
                                                         & \multicolumn{1}{l|}{t2}                                                                            & \multicolumn{1}{l|}{\textbf{Prior (D1), D2}}                                                                & \multicolumn{1}{c|}{\textbf{76.30}} & \multicolumn{1}{c|}{\textbf{79.90}} & \multicolumn{1}{c|}{\textbf{82.73}} & \textbf{92.81} & 2,584 (4)                                                                                     \\ \cline{2-8} 
                                                         & \multicolumn{1}{l|}{t3}                                                                            & \multicolumn{1}{l|}{Prior (D1, D2), D3}                                                                     & \multicolumn{1}{c|}{75.80}          & \multicolumn{1}{c|}{79.80}          & \multicolumn{1}{c|}{82.93}          & 92.76          & 5,104 (8)                                                                                     \\ \cline{2-8} 
                                                         & \multicolumn{1}{l|}{t4}                                                                            & \multicolumn{1}{l|}{Prior (D1, D2), D4}                                                                     & \multicolumn{1}{c|}{75.65}          & \multicolumn{1}{c|}{79.65}          & \multicolumn{1}{c|}{82.83}          & 92.71          & 4,641 (7)                                                                                     \\ \cline{2-8} 
                                                         & \multicolumn{6}{c|}{Sum}                                                                                                                                                                                                                                                                                                                            & \textbf{15,554}                                                                               \\ \hline
\end{tabular}
\end{adjustbox}
\caption{Bayesian/Simple training results on FetaQA}
\label{tab:fetaqa_incremental_training}
\end{table}

\subsection{RQ3. Effect of table representation} 
\label{sec:rq3}

In this section, we demonstrate the impact of \textsf{RCG} by comparing its performance against other
state of the art baselines: 

% \begin{myitemize}

%\noindent$\bullet$ 
\noindent$\bullet$~\textbf{Sliding Token}. Many existing approaches concatenate
the table cells from left to right and include tags to indicate schema
information \cite{yin20acl, chen2021open, herzig2020tapas}. To use this approach
for learned data discovery, the resulting tokens become the input of \textsf{OpenQA},
which we use to obtain a vector embedding. Because the input size of \textsf{OpenQA}
is bounded~\cite{wu2016google}, we tokenize following the approach in \cite{wu2016google} and feed a sliding window over the
cells.
% with size 150 and a stride of size 1. 

%To convert a table to a sequence of tokens, a general approach is to concatenate
%the cells from left to right and add tags to indicate schema information
%\cite{yin20acl, chen2021open, herzig2020tapas}.  We then follow this approach to
%construct a baseline. The TextOpenQA model imposes an upper limit to the input
%passage size in sub-word unit \cite{wu2016google}, which means a word may be
%split into multiple parts by a tokenizer \cite{wu2016google}. It is possible
%that the tokens in a row can exceed that limit so we have to chunk cells.  Often
%neigboring cells can have entity relationships and chunking may cause the
%relationship lost. So we use a sliding window with 150 sub-word  tokens at most
%and use stride of one cell, i.e. the second cell in the previous window is the
%first cell in the next window. Same as RCG, we do it for each row.

%\noindent$\bullet$ 
\noindent$\bullet$~\textbf{RCG\&generated text}. \textsf{RCG} produces (\texttt{subject,
predicate, object}) triples. We ask whether a textual representation of the
triples performs better than the purely structured triples: the intuition is
that text would be a closer representation to the input questions than triples.
In this baseline, we transform triples into text by fine-tuning the T5
model~\cite{raffel2019exploring} using the WebNLG~\cite{gardent2017creating}
dataset. Then, we feed triples to the fine-tuned model and obtain text as
output, which is itself indexed for first-stage retrieval.

% \end{myitemize}

%ones If it can be translated into more fluent text , it
%may help improve the first-stage retrieval.  In addition, the fluent text is
%more close to the text used to train the TextOpenQA model and thus may help
%achieve better input features.  To this end, we finetune T5
%\cite{raffel2019exploring} using theWebNLG \cite{gardent2017creating} dataset to
%translate a triple to natural language text and append it to the triple.  Since
%T5 model is pretrained on huge, diverse and clean text, it can transfer some
%knowlegde to the finetuned modle to help generate more fluent text and possibly
%tokens that can overlap with the question. This will help the retrieval as well.    

\mypar{Experimental Setup} We apply the two baseline strategies to the target
table collection and this results in a vector index and a first-stage result and
a trained relevance model for each  strategy. We show the performance  of both
the first-stage dense index only and the end-to-end system \sys, i.e., index plus
relevance model. 
%\textcolor{blue}{For training data, we use fixed 1,000 sampled questions 
%and train models using Bayesian neural network.}

\begin{figure}[ht]
    \centering
    \includegraphics[width=0.65\columnwidth]{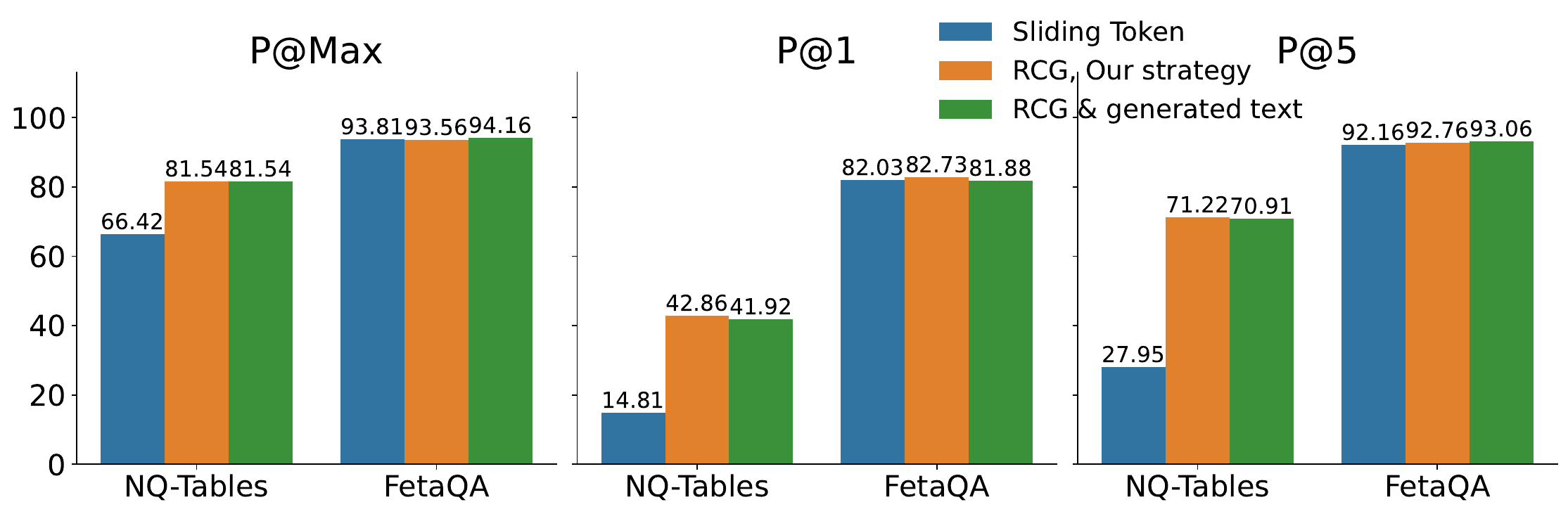}
    \caption{Effect of table representation}
	\label{fig:table_repre_metric}
\end{figure}

\mypar{Results} \F\ref{fig:table_repre_metric} shows the results. \textsf{RCG}
outperforms the other baselines. The plot  shows P@Max (performance
with an oracle second-stage ranking), thus isolating the effect on first-stage
retrieval. While the performance of all baselines is similar on the simpler
\textsf{FetaQA} (cleaner dataset), \textsf{RCG} vastly outperforms the Slide Token baseline in
\textsf{NQ-Tables}. This is because tables in the \textsf{NQ-Tables} baseline have more columns
than \textsf{FetaQA}. 25\% ground truth tables have more than 18 columns with an average of 170
tokens per row. In contrast, in \textsf{FetaQA}, the 75th percentile is 6 columns and
only 17 tokens per row on average. The \textsf{RCG} approach will relate cells no matter
how far apart they are in the table, as per the construction of the complete
graph. In contrast, the common table representation methods from the state of the
art lose context when tables are wide, as evidenced by the Slide Token
performance on \textsf{NQ-Tables}. Finally, the performance gains during first-stage
retrieval carry on to the end-to-end system.

%\begin{figure}[ht]
%\captionsetup[subfloat]{}
%\centering
%\subfloat[Table representation, First-stage retrieval]{{\includegraphics[height=5cm]{img/table_linearization_first_stage_accuracy} }}
%\subfloat[Index type, First-stage retrieval]{{\includegraphics[height=5cm]{img/index_accuracy_first_stage} }}

%\subfloat[Table representation, \sys]{{\includegraphics[height=5cm]{img/table_linearization_accuracy_p_1} }}
%\subfloat[Index type, \sys]{{\includegraphics[height=5cm]{img/index_accuracy_sddnq_p_1} }}

%\subfloat[Table representation, \sys]{{\includegraphics[height=5cm]{img/table_linearization_accuracy_p_5} }}
%\subfloat[Index type, \sys]{{\includegraphics[height=5cm]{img/index_accuracy_sddnq_p_5} }}
%\caption{Effect of table representation and index type}
%\label{fig:table_linearization_accuracy}
%\end{figure}

Finally, generating text from triples (the \textsf{RCG}\&generated text
baseline) makes the system much slower (by requiring an inference from the T5
model per triple) without yielding significant accuracy gains. On further
analysis, we found that the generated text did not really help with
retrieving better content and that in some cases it was erroneous, thus hurting
the end-to-end performance.

\subsection{RQ4. Microbenchmarks}
\label{sec:rq4}

We consider the impact of different indexing techniques on the
first-stage retrieval component (Section~\ref{subsubsec:indexing}) and we
demonstrate the impact of the techniques described in (Section~\ref{subsec:techchallenges}) on runtime in
Section~\ref{subsubsec:runtime}.

\subsubsection{First-Stage Retrieval Indexing.}
\label{subsubsec:indexing}

The first-stage retrieval index determines the input of the relevance model
(Challenge \textbf{C3}), affects the training triples (Challenge \textbf{C1}) and also determines
the table representation (Challenge \textbf{C2}). Here, we measure the effect of choosing
indexing techniques: 

\begin{myitemize}
%\noindent$\bullet$ 
\item \textbf{First Stage (Sparse Index)} We index and retrieve triples using
the BM25 algorithm (on Elasticsearch) that measures the similarity of question
and triple using \textsf{TF-IDF} features \cite{robertson2009probabilistic}. 

%\noindent$\bullet$ 
\item \textbf{First Stage (Dense Index)} This is our first-stage
retrieval implementation which uses the \textsf{Fid} Encoder and \textsf{Faiss} index.

%\noindent$\bullet$ 
\item \textbf{\sys (Sparse Index)} We index triples using
\textsf{Elasticsearch}, and then construct training data from the sparse Index using the
same synthetic questions. We then retrain the relevance model.

\end{myitemize}

%\textcolor{blue}{For training data, we use fixed 1,000 sampled questions 
%and train models using Bayesian neural network.}

\begin{figure}[ht]
    \centering
    \includegraphics[width=0.65\columnwidth]{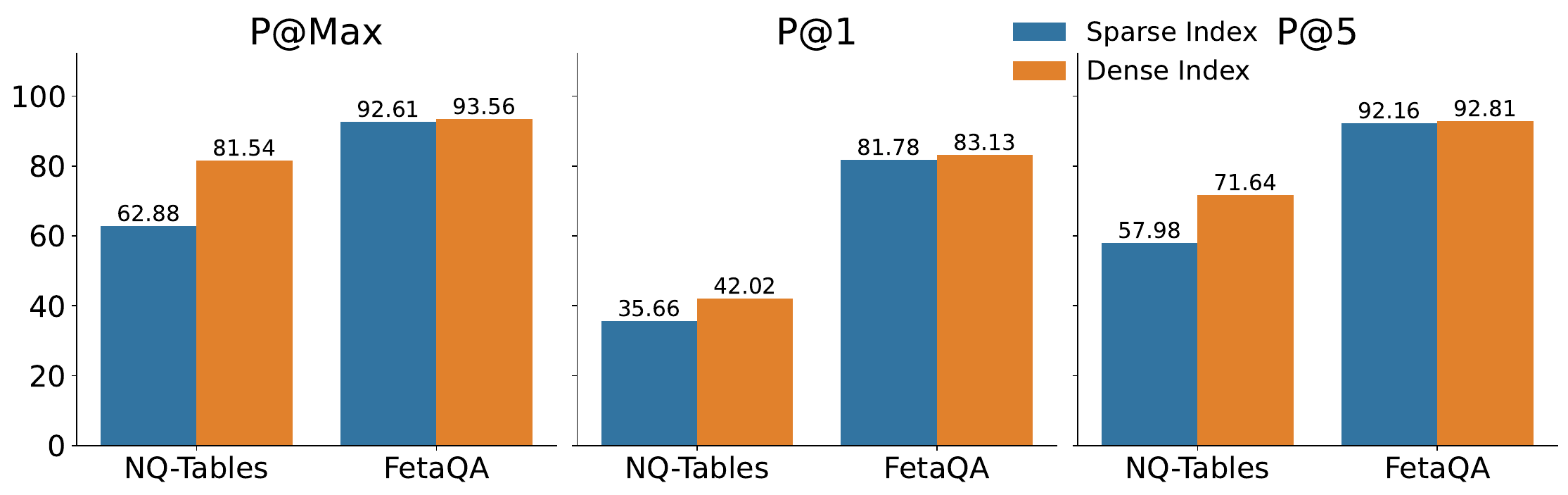}
    \caption{Effect of index type}
	\label{fig:index_type_metric}
\end{figure}

\mypar{Results} \F\ref{fig:index_type_metric} shows the results. The
Dense Index outperforms the two sparse indices baselines. The performance gains
originate during first-stage retrieval, especially for \textsf{NQ-Tables} with
18 points difference and results in 13 points and 6 points in P@5 and P@1.

%Dense index always performs better than sparse index. On
%NQ-Tables, dense index is significantly better than sparse index. The gap is
%clearly originated in the first-stage retrieval and is as large as 17 points
%which results in 12 points above in P@5 and 4 points above in P@1.  This shows
%dense index performs better in retrieving relevant candidate triples by
%comparing in semantic level. 

%\begin{figure}[ht]
%\captionsetup[subfloat]{labelformat=empty}
%\centering
%\subfloat[First-stage retrieval]{{\includegraphics[height=3.3cm]{img/index_accuracy_first_stage} }}
%\subfloat[\sys]{{\includegraphics[height=3.3cm]{img/index_accuracy_sddnq_p@1} }}
%\subfloat[\sys]{{\includegraphics[height=3.3cm]{img/index_accuracy_sddnq_p@5} }}
%\caption{Effect of Index type} 
%\label{fig:index_accuracy} 
%\end{figure}

The Relevance model performs well using the dense index. On
\textsf{FetaQA}, the first-stage gap P@MAX is 0.95 points, but the P@1 gap is
increased to 1.35 points i.e., the second-stage relevance model benefits more
from dense index. The sparse index will retrieve triples based on
the overlap with the input question. Because we produce questions from the
tables and we use the index to construct training data, this indexing
method bias the training dataset towards triples that overlap the most with the
input question. The dense index retrieves triples based on semantic similarity that goes
beyond the purely syntactic level, resulting in higher performance.

%Intuitively, sparse index like BM25 suffers more distribution
%shift between training synthetic and real user questions. To see why, synthetic
%questions unavoidably have more tokens overlapped with triples because questions
%are generated from tables. Thus, in constructing trainining triples, sparse
%index will return triples with more overlaped tokens with the synthetic
%question, and thus the training will bias toward those triples.  While dense
%index retrieves triples in semantic level and thus the bias will be less and the
%triple distribution is more close to that for real test questions, and so dense
%index helps the second-stage relevance model achieve better performance.

%\noindent$\bullet$ More discussion. Using dense index means more time and space
%is needed during offline indexing. But a lot can be done in parallel, e.g.
%split tables into chunks and then encode tables and index the result vectors
%using multiple machines simultenously  and then merge the chunk indexes into
%one. In addition, large-space solid-state drive (SSD) is needed for fast dense
%retrieval and this is also not a big issure with large SSD more and more
%affordable. 

\subsubsection{Run-time Performance.}
\label{subsubsec:runtime}

We evaluate the performance of \textsf{OpenDTR}, \textsf{GTR},  \sys, \sysnostudent (where the teacher encoder is used) and 
\sysnopq (where vectors are added to index without using product quantization). The results on
\textsf{NQ-Tables} are summarized in Table~\ref{tab:pipeline_time}.

% \noindent$\bullet$ Encoding. This step converts tables into numeric vectors.

% \noindent$\bullet$ Building index. This step indexes vectors from Encoding.

% \noindent$\bullet$ Training data collection. This step generates the training
% data.

% \noindent$\bullet$ Training. This step trains the system (model). 

% \noindent$\bullet$ Inference runtime. In this step, we measure the throughput in
% question per second.

\begin{table}[ht]
\resizebox{0.65\columnwidth}{!}{%
\begin{tabular}{|l|c|c|c|c|c|}
\hline
\textbf{Pipeline}                                                                         & \textbf{SoLo}                                                       & \textbf{\begin{tabular}[c]{@{}c@{}}SoLo\\ (No \\ Student)\end{tabular}} & \multicolumn{1}{l|}{\textbf{\begin{tabular}[c]{@{}l@{}}SoLo\\ (No \\ PQ)\end{tabular}}} & \textbf{\begin{tabular}[c]{@{}c@{}}OpenDTR\\ Official\end{tabular}}                 & \textbf{\begin{tabular}[c]{@{}c@{}}GTR\\ Retrain\end{tabular}} \\ \hline
\begin{tabular}[c]{@{}l@{}}Encoding Time \\ (\# vectors)\end{tabular}                      & \begin{tabular}[c]{@{}c@{}}1.15 h\\ (41 M )\end{tabular}             & \begin{tabular}[c]{@{}c@{}}10.95 h\\ (41 M )\end{tabular}                & -                                                                                        & \begin{tabular}[c]{@{}c@{}}0.4 h\\ (0.17M)\end{tabular}                             & -                                                              \\ \hline
\begin{tabular}[c]{@{}l@{}}Building\\ Index Time\\  (Size in GB)\end{tabular}             & \begin{tabular}[c]{@{}c@{}}958 s\\ (16 G)\end{tabular}               & -                                                                        & \begin{tabular}[c]{@{}c@{}}960 s\\ (61G)\end{tabular}                                    & -                                                                                   & -                                                              \\ \hline
\begin{tabular}[c]{@{}l@{}}Training data \\ collection time\end{tabular}                  & 2.4 h                                                                & -                                                                        & -                                                                                        & Manual                                                                              & Manual                                                         \\ \hline
Training time                                                                             & 4.5 h                                                                & -                                                                        & -                                                                                        & 4.6h                                                                                & 4.3h                                                           \\ \hline
\begin{tabular}[c]{@{}l@{}}End-to-endPrediction \\ throughput\\ (Index type)\end{tabular} & \begin{tabular}[c]{@{}c@{}}2.6 s/q\\ (Index \\ on disk)\end{tabular} & -                                                                        & -                                                                                        & \begin{tabular}[c]{@{}c@{}}0.02 s/q\\ (exhaustive search in \\ memory)\end{tabular} & -                                                              \\ \hline
\end{tabular}%
}
\caption{Performance of different systems on NQ-Tables}
\label{tab:pipeline_time}
\end{table}

%\mypar{\textsf{OpenDTR}} We report the numbers from the model released by Google
%We report the training time they declare in their work because we lack resources
%to train such an expensive model.  We reproduce the data for other pipelines
%using this model.
%
%\mypar{\textsf{GTR-Retrain}} This is the the same model as in
%Section ~\ref{sec:rq1}. \textsf{GTR} is not an end-to-end system but a relevance
%model, so we only report the training time.
%
%\mypar{\sys} This is our end-to-end system and each time a
%dataset of size 1,000 is generated as in Section~\ref{sec:rq2}.

%We choose \textsf{NQ-Tables} because these systems achieve
%similar accuarcy on this dataset and it also much larger than the dataset
%\textsf{FetaQA}. 

\mypar{Results} Table~\ref{tab:pipeline_time} shows the results, with one row
for each of the four categories explained above.

% \begin{myitemize}

\noindent$\bullet$~\textbf{Encoding} There are 170k tables in \textsf{NQ-Tables}. 
\sys generates 41M and \textsf{OpenDTR} only 0.17M, one per table. Although \sys produces more vectors than \textsf{OpenDTR}, 
it takes less than 3x time (1.15 h vs 0.4 h) to encode the whole table repository.
\sys is 9.5x (1.15 h vs 10.95 h) faster than \sysnostudent and this makes our system deployment more practical.
% for organizations with large number of tables but sensitive to budget. 

\noindent$\bullet$~\textbf{Building index} \sys supports an on-disk index to scale to larger table collections, 
while \textsf{OpenDTR} uses an in-memory index. It takes only 960 seconds for \sys  to index 41M vectors 
and  \sys uses around 1/4 storage size of \sysnopq. 

\noindent$\bullet$~\textbf{Training data collection} \sys automatically collects training
data and then it takes 2.4 h to train, while both \textsf{OpenDTR} and
\textsf{GTR-Retrained} rely on manual work to collect data. As an example, the \textsf{NQ-Tables} dataset contains 9,534 train questions and 1,067 evaluation questions; collecting these manually is tedious.

\noindent$\bullet$~\textbf{Training runtime} All systems were trained on the same synthetic
dataset generated by our approach so the training time is similar across approaches.
% all systems take a similar amount of time to train.

\noindent$\bullet$~\textbf{Inference throughput} \sys achieves subsecond query latencies, permitting interactive querying. \sys's throughput is lower
(2.6 s/q vs 0.02 s/q) than \textsf{OpenDTR} because \sys does not assume vectors fit in memory, thus working in more general scenarios. If we load vectors into memory, the throughput is similar.
% the table
% vectors do not fit in memory and uses an on-disk index, while
% \textsf{OpenDTR-Official} assumes the table vectors fit in memory. With
% sufficient memory, we expect the throughput of both systems to be similar.

% \end{myitemize}

%\notera{don't understand training runtime}Since \sys automatically determines
%the training data size which is different than the other two using manual data,
%to compare, we convert the training hours  to hours per 1000 data size.  \sys
%and \textsf{GTR-Retrain} have similar training hours per 1000,  while
%\textsf{OpenDTR-Official} is about the half. However \textsf{OpenDTR-Official}
%needs 32 cloud TPU v3, while \sys and \textsf{GTR-Retrain} only use 1 NVIDIA RTX
%6000 GPU and so \textsf{OpenDTR-Official} is still more expensive in training.

The main takeaway message is that the prediction pipeline is slower because of on-disk index search for 
scalability, which we think can be further optimized, the main gain of \sys is its ability to automatically
generate training datasets that lead to high-performance models, as demonstrated
in this section.

\section{Conclusions}
\label{sec:conclusions}

Learned discovery systems offer an attractive way to find data among large table repositories, but they require manual collection of training data, thus limiting their adoption. In response, we introduce a new self-supervised approach to automatically assemble training datasets. We introduce a new table representation method and associated relevance
model that makes the system outperform existing approaches. We build a system, \sys, that implements the new strategies and further includes optimizations to reduce encoding and inference time, and storage footprint. \sys outperforms the state of the art
baselines. All
in all, this work contributes to the growing area of learned
data discovery systems.

\smallskip

%We demonstrate the feasibility of generating training data automatically from a
%large collection of tables for data discovery by natural language question.  We
%show existing state-of-the-art table representation and relevance models are not
%good enough to work with the synthetic training data.  We propose to represent a
%table as a set of (\texttt{subject, predicate, object}) triples, base the
%relevance model on these triples and build an end-to-end self-supervised data
%discovery system. Experiments demonstrate our system can achieve both good
%performance and little human involvement. All in all, our work hints a new
%promising approach to learned data discovery systems.

\bibliographystyle{ACM-Reference-Format}
\balance
\bibliography{main}

\end{document}